\documentclass[12pt]{article}
\usepackage{Xoohm}
\def\USLetter{\topmargin -10 mm
 \oddsidemargin -7 mm \evensidemargin=\oddsidemargin
\textheight 245 mm \hsize 180 mm \textwidth=\hsize}

\USLetter
\newcount\PN \PN=0
\long\def\@#1{\textsf{\global\advance\PN by1?`$^{\the\PN}$[#1]}}
\def\wt#1{\skew6\widetilde#1}
\def\cA{{\cal A}}
\def\cB{{\cal B}}
\def\cH{{\cal H}}
\def\hA{\textit{\textbf{A}}}
\def\ha{\mbox{\boldmath$\alpha$}}
\def\hB{\textit{\textbf{B}}}
\def\hC{\textit{\textbf{C}}}
\def\hD{\textit{\textbf{D}}}
\def\hH{\textit{\textbf{H}}}
\def\hP{\mathbf{\P}}
\def\hp{\textit{\textbf{p}}}
\def\hr{\mbox{\boldmath$\varrho$}}
\def\hs{\hat\sigma}
\def\hS{\textit{\textbf{S}}}
\def\hU{\textit{\textbf{U}}}
\def\hV{\textit{\textbf{V}}}
\def\hW{\textit{\textbf{W}}}
\def\cW{{\cal W}}
\def\hx{\textit{\textbf{x}}}
\def\hz{\textit{\textbf{z}}}
\let\O=\overline

\long\def\Out#1{}
\title{
From Quantum To Classical Dynamics:\\
A Landau Continuous Phase Transition With\\
Spontaneous Superposition Breaking}

\author{Vladan Pankovi\'c$^*$, Tristan H\"ubsch$^\dag$,
        Milan Predojevi\'c$^*$, Miodrag Krmar$^{*, \ddag}$\\[1mm]
\small $^*$Department of Physics, 
        Faculty of Natural Sciences and Mathematics,\\[-2mm]
\small 21\,000 Novi Sad, Trg Dositeja Obradovi\'ca 4.,\\[-2mm]
\small Serbia and Montenegro, {\tt mpgalant@ptt.yu}\\
\small $^\dag$Department of Physics, 
        Howard University, Washington DC 20059,\\[-2mm]
\small {\tt thubsch@howard.edu}\\
\small $^\ddag$Calfiornia State University, 
        Dominguez Hills, 1000 E Victoria Street,\\[-2mm]
\small Carson, California, USA, {\tt mkrmar@csudh.edu}}
\date {}
\begin{document}

\maketitle

\begin{center}
\today
\end{center}

\vspace{5mm}
\begin{abstract}\narrower
Developing an earlier proposal~(Ne'eman, Damnjanovi\'c,\etc), we show herein that there is a Landau continuous phase transition from the exact quantum dynamics to the effectively classical one, occurring via spontaneous superposition breaking (effective hiding), as a special case of the corresponding general formalism~(Bernstein). Critical values of the order parameters for this transition are determined by Heisenberg's indeterminacy relations, change continuously, and are in excellent agreement with the recent and remarkable experiments with Bose condensation. It is also shown that such a phase transition can sucessfully model self-collapse (self-decoherence), as an effective classical phenomenon, on the measurement device. This then induces a relative collapse (relative decoherence) as an effective quantum phenomenon on the measured quantum object by measurement.

We demonstrate this (including the case of Bose-Einstein condensation) in the well-known cases of the Stern-Gerlach spin measurement, Bell's inequality and the recently discussed quantum superposition on a mirror {\it\`a la}  Marshall {\it et al.} These results provide for a proof that quantum mechanics, in distinction to all absolute collapse and hidden-variable theories, is local and objective. There now appear no insuperable obstacles to solving the open problems in quantum theory of measurement and foundation of quantum mechanics, and strictly within the standard quantum-mechanical formalism. Simply put, quantum mechanics is a field theory over the Hilbert space, the classical mechanics characteristics of which emerge through spontaneous superposition breaking.
\end{abstract}
\vfill

\newpage
\Sec{Introduction}
 \Label{s:In}
Extending a supposition from Ref.\cite{R1}, by means of a complex mathematical formalism (superoperator technique) and without an immediate physical explanation, Ref.\cite{1} suggests that there is a Landau continuous phase transition (with spontaneous symmetry breaking)\cite{2} induced by (quantum) measurement. Similar suggestions and indications may also be found in Refs.~\cite{R1,R2,R3,R4,R5}. There, as in Ref.\cite{1}, the underlying symmetry and the mechanism of its breaking is not clarified. In these works, essentially, it is implicitly or explicitly assumed that the spontaneous breaking of symmetry or quantum superposition in every individual measurement occurs under the influence of a finer (quantum-mechanically non-observable) and weak trans-quantum dynamics, the averaging of which over the statistical ensemble ought to reproduce the predictions of the standard quantum-mechanical formalism. That is, such spontaneous symmetry breaking corresponds to what in the general formalism\cite{3,4,5,R4,R5,R6} is called actual or dynamical symmetry breaking, where a small asymmetric dynamical perturbation induces an exact breaking of the unperturbed symmetrical dynamics in every individual case. It is not hard to see that this interpretation directly leads to theories of hidden variables\cite{20}, which are, as is well-known\cite{14,22,23}, necessarily superluminal, \ie, non-local and thus physically unacceptable. (Adopting the concept of non-local theories of hidden variables would necessarily imply discarding not only the theory of relativity, but also of the entire field theory as we know it\cite{5}.)
 
 Herein, the identification of measurement with a spontaneous symmetry (superposition) breaking (effective hiding) is formalized and provided with a clear physical explaination. That is, we will show that there is a typical Landau continuous phase transition from quantum to classical dynamics, with a corresponding spontaneous breaking (effective hiding) of the quantum superposition, \ie, the global quantum-dynamical $U(1)$ symmetry. Critical values of the order parameters for this phase transition are determined by Heisenberg's indeterminacy relations, change continuously, and are well supported by recent experiments with Bose condensation\cite{ref.1,ref.2}. This spontaneous superposition breaking represents a particular example of the general formalism of spontaneous symmetry breaking (effective hiding)\cite{3,4,5,R4,R5,R6}, applicable in diverse domains of physics (classical mechanics of deformable bodies, quantum theory of ferromagnetism, quantum theory of unified fields, \etc). Just as all these applications of symmetry breaking correspond to a phase transition, so too does the present case: it is a transition from the (exact) quantum dynamics to the (approximate) classical dynamics (obtained, in turn, by standard averaging of the quantum dynamics). And, while superposition and the global $U(1)$ symmetry remain unbroken in the quantum analysis, classical dynamics breaks them.
 
 This phase transition can be used to model the self-collapse (self-decoherence) on the measurement device which, together with a correlative dynamical interaction between the measurement device and the measured quantum object, induces the relative collapse (relative decoherence) as an effective quantum phenomenon effected on the measured quantum object by the (quantum) measurement. This provides for solving all open problems in quantum theory of measurement and fundation of the quantum mechanics\cite{6,7}, and strictly within standard quantum-mechanical formalism\cite{6,8,9,10,11}. We illustrate this by the example of the Stern-Gerlach's spin measurement\cite{12,13}, Bell's inequality\cite{14} (including the proof that quantum mechanics, in distinction to all absolute collapse\cite{6}, non-unitary (nonlinear) dynamics\cite{15}, absolute environmental decoherence\cite{16,17,18,19} and hidden variable theories\cite{20}, is local and objective), and the quantum superposition on a mirror experiment, recently suggested by Marshall {\it et al.}\cite{21}.
 
 Simply put, instead of the widely accepted belief (characteristic of theories with absolute collapse and hidden variables) that quantum mechanics is an essentially mechanistic theory (a discretuum theory with actions at a distance), \ie, that through a convenient extension of the standard formalism (through nonlinear terms, \etc) turns quantum mechanics into a mechanistic theory, we prove the opposite: Quantum mechanics is in fact a true local field theory over a Hilbert space (see appendix~\ref{s:SSS}), the classical mechanics characteristics of which emerge through a spontaneous superposition breaking (see Sec.~\ref{s:SQS},~\ref{s:QSS}, and appendices~\ref{s:PROB} and~\ref{s:WIWP}), such as it happens, \eg, in the measurement process (see Sec.~\ref{s:MeaSSB}). This will, entirely within the standard quantum-mechanical formalism, establish a natural relationship between the general formalism of classical and quantum mechanics; in fact, this also provides a relationship between quantum mechanics and quantum field theory.

\Sec{A Simple Quantum System}
 \Label{s:SQS}
Consider a simple quantum system with the unitary quantum-dynamical evolution operator, $\hU(t)$, and a quantum-dynamical state of unit norm\ft{Throughout, operators are set in boldface type, and vector spaces in script.}:
\begin{equation}
 \ket{\J\6(t)}=\hU(t)\ket{\J} = \sum_n c_n \ket{u_n\6(t)}~,
\Label{e:1}
\end{equation}
which evolves deterministically in time, $t$. Here, $\ket{\J}$ represents the initial quantum state, while $c_n$ for $\forall n$ represent the constant superposition coefficients, which satisfy the unit norm condition:
\begin{equation}
 \sum_n |c_n|^2 = 1~. 
\Label{e:2}
\end{equation}
Writing $\cB =\{\ket{u_n}, \forall n\}$ for the initial basis in a Hilbert space, $\cH_{qp}$, on which the action of coordinate and momenta observables (and analytic functions thereof) are defined\ft{This Hilbert (sub)space is sometimes referred to as the ``orbital Hilbert space'', in a direct reference to the Hydrogen atom, and in contradistinction from the Hilbert (sub)space spanned by the spin factors in that archetypical case. More generally, a distinction between ``positional'' and ``orientational'' variables--and hence subspaces of the total Hilbert space--might be preferable. Aiming here at a general setting, we avoid this reference, supplant it with the arguably more verbose but also more general characterization, and use ``$\cH_{qp}$'' for this ``positional Hilbert (sub)space'' hereafter.},
\begin{equation}
\cB(t)=\hU(t)\cB=
  \{\hU(t)\ket{u_n}=\ket{u_n\6(t)}, \forall n\}
\Label{e:2a}
\end{equation}
then represents the corresponding time-dependent basis.
 It is well known that quantum dynamics~(\ref{e:1}) is \textit{globally} $U(1)$-symmetric, \ie, that it is invariant with respect to the following \textit{global} transformation
\begin{equation}
 \ket{\J\6(t)} \mapsto \ket{\J'\6(t)}= e^{i \a}\ket{\J\6(t)}~,\qquad
  \forall t,~ \forall\J~,
\Label{e:3}
\end{equation}
where $\a$ is an arbitrary real constant. 

\sSec{Symmetry and superposition breaking}
 \Label{s:SSB}
We next define the following unitary operator:
\begin{equation}
 \hW_\ve[\J\6(t)]=\exp\{i \ve \hP_\J(t)\}~,\qquad
 \textrm{where}~~\hP_\J(t)=\ket{\J\6(t)}\bra{\J\6(t)}~,
\Label{e:4}
\end{equation}
with $\ve$ a real parameter and the projector $\hP_\J(t)$, which generates the transformation. Obviously, we then have that:
\begin{equation}
 \hW_\ve[\J\6(t)]\,\ket{\J\6(t)}=e^{i\ve}\ket{\J\6(t)}~,
\Label{e:5}
\end{equation}
and, on the state $\ket{\J\6(t)}$, $\hW_\ve(\J(t))$ reproduces the transformation~(\ref{e:3}).
Since this leaves the quantum dynamics invariant, the state~(\ref{e:5}) is equivalent to~(\ref{e:1}). Therefore, as a symmetry of the particular quantum state $\ket{\J\6(t)}$ (and so ``ultra-local'' on the Hilbert space; see appendix~\ref{s:SSS} for a more precise definition), $\hW_\ve[\J\6(t)]$ represents a subgroup of the global quantum-dynamical $U(1)$ symmetry.
 It is not hard to see that this quantum-dynamical state symmetry, $\hW_\ve[\J\6(t)]$, conserves exactly the quantum dynamical state $\ket{\J\6(t)}$ as a superposition, with constant coefficients, in any time dependent basis of $\cH_{pq}$, which evolves equivalently to $\ket{\J\6(t)}$, \ie, according to the quantum-dynamical evolution of the system. In this sense, $\hW_\ve[\J\6(t)]$ prefers none of these bases.

Now suppose that there exists a time $\tau$, so that for $t>\t$
the following two approximation conditions are satisfied: 
\begin{itemize}
 \item[$A_1$:] Any quantum state from $\cB(t)$ represents a wave packet. As well known\cite{9}, then (except, of course, for any one fixed $k$ and $\hA_k\define\hA{-}\vev{u_k|\hA|u_k}$):
\begin{equation}
 \Big|\bra{u_n\6(t)}\hA\ket{u_n\6(t)}\Big| \gg
 \triangle_{u_n\6(t)}\hA\equiv
 \sqrt{\vev{u_n\6(t)|\hA^2|u_n\6(t)}-\vev{u_n\6(t)|\hA|u_n\6(t)}^2},\qquad
  \forall n ~,
\Label{e:7}
\end{equation}
for any observable $\hA$ defined to act on $\cH_{qp}$. (All such observables may be given as formal analytical functions of the momentum or coordinate observables, defined to act on $\cH_{qp}$.)
 \item[$A_2$:] All quantum states from $\cB(t)$ are weakly interfering. That is,
\begin{equation}
 \Big|\vev{u_n\6(t)|\hA|u_m\6(t)}\Big| \simeq
 \Big|\vev{u_n\6(t)|\hA|u_n\6(t)}\Big| \delta_{nm}~,\qquad \forall n, m~,
\Label{e:8}
\end{equation}
or, equivalently:
\begin{equation}
 \Big|\vev{u_n\6(t)|\hA|u_n\6(t)} - \vev{u_m\6(t)|\hA|u_m\6(t)}\Big| \geq
 \inv2 (\triangle_{u_n\6(t)}\hA + \triangle_{u_m\6(t)}\hA),\qquad
  \forall n, m\neq n~,
\Label{e:9}
\end{equation}
for any observable $\hA$ defined over $\cH _{qp}$.
\end{itemize}
It is very important to note that, according to the standard quantum-mechanical formalism\cite{6,8,9}, there is an up to the global $U(1)$-equivalence~(\ref{e:3}) uniquely determined basis, $\cB(t)$, which satisfies both approximation conditions, $A_1$ and $A_2$, after a time $\tau$. 

Subject to these approximation conditions $A_1$ and $A_2$, Eqs.~(\ref{e:4}) and~(\ref{e:5}) turn into
\begin{equation}
 \wt\hW_\ve[\J\6(t)] \simeq\exp\{i \ve \sum_n |c_n|^2\hP_{u_n}(t)\}~,\qquad\mbox{and}
\Label{e:10}
\end{equation}
\begin{equation}
 \wt\hW_\ve[\J\6(t)]\,\ket{\J\6(t)}
   \simeq\sum_n c_n\exp\{i\ve|c_n|^2\}\ket{u_n\6(t)}~,
\Label{e:11}
\end{equation}
where the cross-terms in $\hP_\J=(\sum_n c_n\ket{u_n\6(t)})(\sum_{n'} c^*_{n'}\bra{u_{n'}\6(t)})\simeq\sum_n |c_n|^2\hP_{u_n}(t)$ are being neglected.

It should be clear that the resulting, transformed quantum state~(\ref{e:11}) is not a globally $U(1)$-equivalent copy of the quantum-dynamical state~(\ref{e:1}). That is, in general, the various superposition terms in~(\ref{e:11}) clearly acquire different phases, so that they, relative to the superposition~(\ref{e:1}), become mutually decoherent. It should be clear that this decoherence is very similar to the decoherence between classical waves, and rather dissimilar from the absolute environmental decoherence in Refs.~\cite{16,17,18,19}. This then implies that the approximation~$A_1,A_2$ breaks both the quantum-dynamical state symmetry~(\ref{e:5}), and also the global quantum-dynamical $U(1)$ symmetry~(\ref{e:3}). That is, the result of the approximation~$A_1,A_2$ cannot be achieved by a quantum-dynamical evolution: it is discontinuously inaccessible by exact quantum-dynamical evolution.

It is well known\cite{8,9} that only those quantum-dynamical states which satisfy the wave packet approximation~$A_1,A_2$ can be consistently regarded as classical dynamical states, \ie, as classical particles. In other words, a quantum-dynamical state can be stable in the sense of classical dynamics\ft{We suppose, and show (see appendix~\ref{s:WIWP}), that classical dynamics represents a well-defined theory, specified as a simple (by quantum averaging only) approximation of the exact quantum dynamics.} only if it satisfies the wave packet approximation. And, the other way around, a quantum-dynamical state that does not satisfy the wave packet approximation cannot be classically stable. Furthermore, classical dynamics is indeed a well-defined, approximating limit of quantum dynamics, obtained by \textit{standard} averaging; see below (and especially appendix~\ref{s:WIWP}) for more details.

Now, it is not hard to see from Eqs.~(\ref{e:7}) and~(\ref{e:8}), that no nontrivial (quantum) superposition of weakly interfering wave packets can itself represent a wave packet. This then implies that every quantum-dynamical superposition of weakly interfering wave packets, while quantum-dynamically stable, must be classically unstable. Nevertheless, any quantum superposition of weakly interfering wave packets contains \textit{locally} stable terms in the sense of classical dynamics, corresponding to any wave packet from the given superposition. For this reason, any superposition of weakly interfering wave packets is a perfectly stable state at the quantum level of the analysis. However, in the approximate classical analysis, the same superposition necessarily turns spontaneously and arbitrarily into one of its locally stable terms, \ie, one of the wave packets.

Thus, the weakly interfering wave packet approximation, $A_1$, $A_2$, induces a spontaneous ``transition'' of the quantum superposition, $\ket{\J\6(t)}$, into one of its constituent wave packets, $\ket{u_n\6(t)}$, with the well-known probability (see appendix~\ref{s:PROB})
\begin{equation}
 w_n = |c_n|^2 ~,
\Label{e:12}
\end{equation}
but only at the approximate, classical level of the analysis\ft{In lieu of a standard term and for want of a better word, we will refer to this transition as ``classicizing.''}. (In the exact, quantum dynamics there is no transition from $\ket{\J\6(t)}$ at this or any other fixed moment $t>\t$.) This transition represents a spontaneous quantum superposition breaking by the classical analysis. More generally, we observe a spontaneous breaking (effective hiding) of the global $U(1)$ quantum-dynamical symmetry and the $\hW_\ve[\J\6(t)]$ quantum-dynamical state symmetry at the approximate, classical level of analysis and in the uniquely determined basis, $\cB(t)$.

Here we refer to the general definition of spontaneous symmetry breaking, wherein the observed state (classical particle) lacks the symmetry of the (exact, quantum-) dynamical equations owing to the imposition of ``boundary conditions'' (the wave packet approximation: $A_1,A_2$).

Indeed, the formal similarity of this spontaneous and arbitrary choice of one of the constituent wave packet states, $\ket{u_n\6(t)}$, with the arbitrary choice of a direction in which an originally axially symmetric rod bends under longitudinal compression is inescapable. Other examples of general formalism of the spontaneous symmetry breaking (effective hiding at an approximate level of the analysis)\cite{3,4,5} readily come to mind, such as the spontaneously and arbitrarily chosen orientation of ice crystals emerging in super-cooled water, or of magnetic domains in a ferromagnet. The defining features of spontaneous symmetry breaking are present in all of them.

\sSec{Some details of the spontaneous superposition breaking}
 \Label{s:DetSSB}
Besides the above quoted very essence of spontaneous symmetry breaking, we note, more precisely, that in any such situation there are two coexisting, but radically distinct types of solutions of the dynamical equations: those with exact (unbroken, explicit, complete) symmetry, and, those with approximate (broken, hidden, reduced) symmetry. The former preserve the original superposition, while the latter break it. Either of these can be chosen quite arbitrarily, but a formal transition from the symmetric state into a state with broken symmetry is not describable by the underlying exact (quantum) dynamics.

To see this, note that the action of the exact (quantum) evolution operator, $\hU(t)$, is linear~(\ref{e:1}), and hence by definition preserves superposition. Therefore, this evolution operator cannot bring about the superposition breaking seen as a distinction between the exact (quantum) solutions and the approximate (classical) ones.

In our present situation, the cause for this transition is provided by the imposition of the wave packet approximation, corresponding to the approximate, classical level of analysis. (This then additionally corresponds the approximateness of the classical level of analysis to the approximateness of its solutions.) Furthermore, viewing the formal transition from $\ket{\J\6(t)}$ to one of its wave packet constituents, $\ket{u_n\6(t)}$, at the approximate, classical level of analysis, this spontaneous superposition breaking provides a \textit{classical} stabilization of the system. (At the exact, quantum-dynamical level, such stabilization is neither needed nor does it occur.)

Spontaneous symmetry breaking in general has the well-known consequence of the appearance of Goldstone bosons. For a complete identification of the above process as spontaneous symmetry breaking, we must turn to identifying this feature as it is manifested in our case. To this end, note that Eq.~(\ref{e:12}) quotes the \textit{a priori} probability for the transition $\ket{\J\6(t)}\to\ket{u_n\6(t)}$. \textit{A posteriori}, the probabilities become $\tilde{w}_k=\d_{nk}$, $\forall k$, with $n$ fixed by the given transition. The Goldstone mode would then have to correspond to a transformation involving all of the components, $\ket{u_k\6(t)}$, of the original superposition, $\ket{\J\6(t)}$---for all but one of which, however, the \textit{a posteriori} probability vanishes. This in turn makes the Goldstone mode(s) unobservable in all classical (approximate) systems of analysis, in each of which an arbitrary but definite $\ket{u_n}$ has been spontaneously selected. Amusingly, in this respect, the quantum analysis agrees: Being exact, it preserves the superposition, breaks (hides) no symmetry, and so induces no Goldstone mode. A more detailed analysis of the involved symmetry structure and breaking is presented in Appendix~\ref{s:SSS}.

Finally, it is not hard to see that right-hand side of Eq.~(\ref{e:9}) is practically a constant, which is, through $A_1$, determined by Heisenberg's indeterminacy relations. On the other hand, the left-hand side of Eq.~(\ref{e:9}) is time-dependent. Thereby, Eq.~(\ref{e:9}) becomes the condition for the phase transition\cite{2}, where the left-hand side of Eq.~(\ref{e:9}) plays the r\^ole of the order parameter, continuously changing in time, while the right-side of Eq.~(\ref{e:9}) represents the critical value of this order parameter, whence this corresponds to a Landau , continuous phase transition. 

\sSec{Spontaneous superposition breaking and experiments}
 \Label{s:ExSSB}
So far, we have shown that the use of a weakly interfering wave packet basis spontaneously breaks (hides) superposition of such states and with it the (ultra-local in the Hilbert space) symmetry~(\ref{e:5}). It of course remains to explore if this spontaneous superposition breaking is experimentally verifiable.

The experimental preparation of a simple quantum system into a wave packet (a particle for a ``macroscopic'' classical model) during a finite time period (during which the dissipation of this packet may be neglected) is rather simple. However, the preparation of a quantum system into a superposition of weakly interfering wave packets, if possible at all, appears to be technically very complicated so that it has thus far not been realized. It is thus not clear whether such a superposition of ``classical particles'' can be achieved in practice, nor whether its breakdown occurs explicitly, dynamically or spontaneously.

Nevertheless, we should like to argue that an experimental verification of spontaneous breaking (hiding) of superposition on a simple quantum system is possible, in agreement with the above. In fact, it is possible to so interpret the recent experimental results of Refs.~\cite{ref.1,ref.2}, confirming Bose-Einstein condensation~\cite{ref.3}.

In an ideal gas of Bose quantum systems (in real experiments, these are atoms of Rubidium~\cite{ref.1}, Sodium~\cite{ref.2}, \etc) in thermodynamic equilibrium at temperature $T$  with its environment, the thermodynamical average value of momentum of each individual system is $\vev{p}=\sqrt{mk_{\sss B}T}$, where $m$ is the mass of the system and $k_{\sss B}$ the Boltzman constnat. According to de Broglie's relation, the thermodynamical average value of the corresponding wavelength is then $\vev{\l}=h/\vev{p}=h/\sqrt{mk_{\sss B}T}$, whereby $\vev{\l}$ decreases with the temperature $T$ and \textit{vice versa}.

   Roughly\cite{ref.1,ref.2,ref.4}, if $\vev{\l}$ is less than the thermodynamical average distance between two Bose systems, $\vev{\triangle x}$, and which holds for temperatures above a certain critical value, $T_c$ (in real experiments about $10^{-9}$\,K, which is far below common everyday temperatures), then the wave packets of these systems are indeed approximately weakly interfering, and the systems are effectively both separated and separable (may be identified individually) so that they obey Boltzman statistics. However, if $\vev{\l}>\vev{\triangle x}$, that is, for $T<T_c$, the wave functions of the individual Bose systems no longer satisfy the wave packet approximation and also do interfere. The Bose systems are no longer resolvable and they form a single Bose-Einstein super-system (condensate, collective), which can no longer be described by the wave packet approximation of the quantum state, and which itself obeys the Bose-Einstein statistics.

    According to Einstein's original considerations\cite{ref.3}, the formation of Bose-Einstein condensate (for $T<T_c$) occurs strictly statistically, that is, totally spontaneously and without any (non-ideal, additional, dynamical) interaction between the Bose (sub)systems. Also, the de-condensation\ft{Strictly speaking, \textit{evaporation} refers to a transition into the gaseous phase only near the surface of the liquid and \textit{sublimation} is the analogue for a solid. As the phase transition is here occurring throughout the volume of the substance, \textit{boiling} would perhaps be a more adequate term, except that this also carries connotations of bubbling, turbulence and dynamical vehemence. As neither of these need apply in the case of the reverse of Bose-condensation, we will adhere to the less richly associated term \textit{de-condensation}.} of the condensate (for $T>T_c$) occurs totally spontaneously, without any interaction between the Bose (sub)systems.

    Note that, in real experiments\cite{ref.1,ref.2}, it is impossible to completely suppress the (non-ideal, dynamical) interactions between the alkali atoms (as model Bose systems), or the interaction between these and the environment. These additional interactions affect the instability of the Bose condensate, and determine the mode of its decay even while $T<T_c$. Thereby the alkali atoms cannot serve as an effective model for an ideal gas of Bose systems for very long times. Nevertheless, during sufficiently short periods (a few seconds) during which an effective approximate modeling of an ideal gas of Bose systems by alkali atoms is possible, the said additional interactions have no influence on the spontaneous character of forming (for $T<T_c$) or decaying (for $T>T_c$) of the Bose condensate. Furthermore, it is not hard to see that assuming a sizable effect of these additional interactions on the  statistics of the ideal Bose gas would lead to irreconcilable contradictions with the existing experiments.

     Thus we conclude that, during the time period when it is possible to effectively model an ideal Bose gas using alkali atoms, the experimentally verified Bose condensation (as well as de-condensation) represents a typical Landau continuous phase transition, where the temperature, $T$, may be regarded as the  continuously variable order parameter, with the critical value $T_c$.

      On the other hand, following the above arguments, Bose condensation occurs when the approximation of weak interference between the wave packets begins to fail, and the exact superposition becomes notable. Similarly, Bose de-condensation occurs when this approximation becomes valid and the superposition becomes hidden. In addition, $T_c$ is determined by the characteristics of the wave packet approximation, \ie, Heisenberg's indeterminacy relations. (``The Bose-Einstein condensate therefore is a rare example of the uncertainty principle in action in macroscopic world.''\cite{ref.4}) Thus, it necessarily follows that the Bose de-condensation is a special case of spontaneous superposition breaking (hiding), just as Bose condensation is a special case of spontaneous revealing of superposition. In this precise sense, the spontaneous superposition breaking (hiding) in a simple quantum system described theoretically in \SS~\ref{s:SSB}--\ref{s:DetSSB} is an experimentally verified phenomenon.

 It remains to clarify the conceptual conundrum stemming from a possible inference from the above discussion that the Bose condensate was treated as a simple quantum system (without separable sub-systems) although it is, by common intuition, understood as composed of a collection of individual Bose (sub)systems, such as alkali atoms. Conversely, it remains to dispel the apparent contradiction between the quoted and experimentally verified statement that the quantum states of different Bose systems may be superposed and do interfere\cite{ref.1,ref.2}, on one hand, and the fact that a superposition of quantum states of different quantum systems is impossible within the standard quantum-mechanical formalism\cite{6,8,9,10,11}, on the other.
    
 Suppose that, $1,2,{\cdots},n$ label simple quantum systems (themselves containing no sub-systems), which are exactly described (in Schr\"{o}dinger's picture, say) by the states $\ket{\J_1},\ket{\J_2},{\cdots},\ket{\J_n}$, belonging to the Hilbert spaces $\cH_1,\cH_2,{\cdots},\cH_n$, respectively. It is then possible to define the quantum super-system $1{+}2{+}{\cdots}{+}n$ with sub-systems $1,2,{\cdots},n$, exactly described by the quantum state $\ket{\J_1}{\otimes}\ket{\J_2}{\otimes}{\cdots}{\otimes}\ket{\J_n}$, from the Hilbert space $\cH_1{\otimes}\cH_2{\otimes}{\cdots}{\otimes}\cH_n$, where $\otimes$ is the tensor product. This quantum state of the super-system is called non-correlated or non-entangled. However, there exist also so-called correlated or entangled quantum states of the super-system which are non-trivial superpositions of the non-correlated sub-systemic states.
    
 Following the standard quantum-mechanical formalism\cite{6,8,9,10,11,14} and the empirical results\cite{22,23}, the super-system $1{+}2{+}{\cdots}{+}n$ can be regarded as comprised of the sub-systems $1,2,{\cdots},n$ if and only if the so defined super-system is described by a non-correlated (non-entangled) quantum state. Conversely, if the super-system $1{+}2{+}{\cdots}{+}n$ is described by a correlated (entangled) quantum state---it must be regarded as a simple system, inseparable into the sub-systems $1,2,{\cdots},n$. This implicitly indicates the possibility that the quantum state in a Hilbert space represents an intrinsic ontology of the quantum system and not just an abstract construction.
    
   Suppose now that $1,2,{\cdots},n$ represent Bose systems of an ideal gas. Let $\cB\define\{\ket{\J_k},\>\forall k\}$ be an eigen-basis of the observable of energy (and linear momentum, in this special case), of non-degenerate spectrum in the Hilbert space $\cH$ for $n=1$ (where we may drop the index 1). For some $n$ quantum states taken from $\cB$, $\{\ket{\J_{k_1}},\ket{\J_{k_2}},{\cdots}\ket{\J_{k_n}}\}$, the Bose super-system may be exactly described by the quantum state:
\begin{equation}
 \ket{\J^{1+2+\cdots+n}}\define\inv{\sqrt{n!}}\sum_J
     \hs_{\!_J} \ket{\J^1_{k_1}}{\otimes}\ket{\J^2_{k_2}}{\otimes}{\cdots}
            {\otimes}\ket{\J^n_{k_n}}~,
\end{equation}
where $\hs_{\!_J}$ performs the $J^{th}$ permutation of subscripts of the uncorrelated quantum state on which it acts, so $J=1,2,{\cdots},n!$. Clearly, $\ket{\J^{1+2+\cdots+n}}$ is a nontrivial superposition of non-correlated quantum states, and so is itself a correlated quantum state. This then implies that the exact quantum-mechanical Bose super-system $1{+}2{+}{\cdots}{+}n$ must be regarded as a simple quantum system, and that here $1,2,{\cdots},n$ cannot correspond to realistically separable Bose sub-systems of the super-system; here, $1,2,{\cdots},n$ may be spoken of as sub-systems only conditionally, \ie, formally. Thus, the correlated quantum state, \ie, superposition $\ket{\J^{1+2+\cdots+n}}$ precisely determines the concept of a Bose collective, condensate or super-atom.
    
    However, if the condition that the quantum states from $\cB$ may be regarded, approximately, as weakly interfering wave packets is satisfied, the correlated quantum state of the super-system, \ie, the superposition $\ket{\J^{1+2+\cdots+n}}$ spontaneously decays into its individual superposition terms. That is, $\ket{\J^{1+2+\cdots+n}}$ undergoes a phase transition into a mixture of uncorrelated quantum states
 $\hs_{\!_J}\ket{\J^1_{k_1}}{\otimes}\ket{\J^2_{k_2}}{\otimes}{\cdots} {\otimes}\ket{\J^n_{k_n}}$
 for $J=1,2,{\cdots},n!$, and with equal probabilities, \ie, statistical weights of $\inv{n!}$ each. This then, according to the principles of the standard quantum-mechanical formalism, also means that the Bose super-system (condensate, collective, super-atom) $1{+}2{+}{\cdots}{+}n$ really decays into separable sub-systems $1,2,{\cdots},n$.
    
 Note that the energy of each non-correlated state in the mixture is perfectly equal, \ie, that the foregoing assumptions imply that all non-correlated quantum states of the super-system have identical expressions for the energy of the super-system. This permits the further approximation in which one strictly accounts only for the distribution of the energy of the super-system across the sub-systems but not for their spatial position. This approximation is characteristic of statistical mechanics and thermodynamics.
    
 We may thus define the thermodynamic averaging for large $n$ and equilibrium processes characterized by the temperature $T$. In addition, we assume that all dynamical interactions of the Bose super-system $1{+}2{+}{\cdots}{+}n$ and its environment may be effectively described as the action of a field on the super-system and not a correlated dynamical interaction of the super-system and the environment. The thermodynamic average quantum state of the super-system may be represented in the form of a correlated quantum state
\begin{equation}
 \O{\ket{\J^{1+2+\cdots+n}}}
~=~\inv{\sqrt{n!}}\sum_J \hs_{\!_J}\O{\ket{\J^1_{k_1}}}{\otimes}
  \O{\ket{\J^2_{k_2}}}{\otimes}{\cdots}{\otimes}\O{\ket{\J^n_{k_n}}}~,
\end{equation}
where $\O{\ket{\J^i_{k_i}}}$ denote the thermodynamically, \ie, statistically averaged (energetically most favorable) quantum states on the $i^{th}$ Bose sub-system. To these thermodynamically favorable quantum states of the sub-system we may, in the wave packet approximation, ascribe thermodynamic, \ie, statistically averaged (most favorable) de~Broglie wave lengths $\O{\l}_{k_1},\O{\l}_{k_2},{\cdots},\O{\l}_{k_n}$.
    
 Even with no further details, it follows that for $T>T_c$, \ie, in the case of weak interaction of the thermodynamically averaged sub-system states as wave packets, the thermodynamically averaged correlated quantum state (the Bose super-system superposition) spontaneously decays into a mixture of non-correlated super-system states, and each such non-correlated state is represented by a (factorized, \ie, separated) tensor product of thermodynamically averaged sub-system quantum states. Conversely, for $T<T_c$, \ie, upon failure of the weak interaction approximation for the thermodynamically averaged sub-systems states, the resulting mixture of non-correlated super-system states spontaneously passes into a thermodynamically averaged correlated quantum state, \ie, into a super-system superposition.
    
 This, we hope, clarifies the above conceptual conundrum.

   In much the same way, some other theoretical predictions\cite{ref.5} and their experimental confirmations\cite{ref.6} about the de Broglie wavelengths of two- and multi-photon wave packets, \ie, correlated photon super-systems, may also be regarded as an indirect confirmation of the existence of spontaneous superposition breaking.

\sSec{A topical summary}
 \Label{s:Sum}
Let us indulge in a brief and heuristic review of the properties uncovered thus far, prompting our identification of the superposition and symmetry breaking, and comparing with some well known examples of this ubiquitous phenomenon.

\ssSec{Underlying microphysics}
In the well-known (even everyday) examples of phase transition, the microscopic physics is understood to govern all important dynamical aspects. Thus, for example, in the case of a thin, straight rod compressed longitudinally on both ends it is these boundary conditions that predetermine the ultimate geometry of the bending of the rod. In much the same way and as discussed above, it is the details of the weakly interfering wave packet dynamics (required for a classical dynamics) that predetermines the decoherence phenomenon.

\ssSec{Arbitrary selection}
The actual direction (with respect to the laboratory, Earth, Universe,\ldots) in which the rod eventually bends is (in the absence of impurities, inhomogeneities and transversal external influence) totally arbitrary. In the same vein, the choice of one of the (very) many weakly interfering wave packets onto which the decoherence focuses is totally arbitrary and random.

\ssSec{Ordering parameter}
The rod bends owing to the fact that the externally imposed compressive forces overcome the deformability parameters of the material; this of course identifies the compressive force as the ordering parameter. In our case, of a ``classicization'' of an inherently quantum dynamics, the ordering parameter is the ``distance'' between the expectation value of the given operator (characteristic for the given process) in one and the other weakly interfering wave packet state, \ie, in one and the other classiciziation. Thus, once two possible classicizations (pointilizations) begin to differ substantially, the quantum dynamics becomes effectively described by the classical.

\ssSec{Symmetry breaking}
In the case of the bending of an initially perfectly straight rod, the continuous (rotational, axial $U(1)$) symmetry of the rod breaks down to nothing. (More generally, one expects a possibly nontrivial subgroup of the original symmetry group.) As discussed in \SS~\ref{s:SSB}--\ref{s:DetSSB}, and in more detail in appendix~\ref{s:SSS}, in the case of the (phase) transition from a general superposition, $\ket{\J}$ to a constituent weakly interacting wave packet, $\ket{u_n}$, the symmetry which is being broken is generated by the projector $\hP_\J(t)\define\ket{\J\6(t)}\bra{\J\6(t)}$ and is closely related to the quantum-dynamical $U(1)$ symmetry~(\ref{e:5}).

\ssSec{Goldstone modes...}
 \Label{s:GM}
It is a well-known (and rigorous) theorem that all symmetry breaking-physical processes induce Goldstone modes. In the case of the bending rod, this is seen as follows: before bending, the rod has two transversal vibrational modes, both of which have nonvanishing frequencies/energies; after the bending, the radial vibrational mode still has a nonzero frequency/energy, but the rotational mode (about the axis defined by the previously unbent rod) has a vanishing frequency: this is the Goldstone mode. Note that, as a (previous symmetry) \textit{transformation}, the Goldstone mode represents a transition from one of the possible results (directions) of the bending of the rod into another, and ranges over all of them.

In the case of classicization of quantum dynamics by weakly interfering wave packets, the Goldstone mode must correspond to a transformation amongst all of the various components, $\ket{u_k\6(t)}$, of the original superposition, $\ket{\J\6(t)}$. It is therefore generated by operators including all those of the type $\ket{u_i\6(t)}\bra{u_j\6(t)}$, $\forall i\neq j$, ``rotating'' any one of the possible classicizations of the inherently quantum dynamics into another, and ranging over all possibilities; see appendix~\ref{s:SSS}.

\ssSec{...are unobservable}
 \Label{s:NoGM}
However, \SS\,\ref{s:DetSSB} shows that the \textit{a posteriori} probabilities for the transition into any of the weakly interfering wave packets are $\tilde{w}_k=\d_{nk}$, $\forall k$, so that the probability into any one of the not realized ones is $\tilde{w}_k=0$, $k\neq n$. Therefore, not one of the possible Goldstone modes represents an observable within any one of the realized classicizations of the quantum dynamics\ft{We couldn't help but notice the potential of an amusing application of such Goldstone transitions as a Sci-Fi vehicle of travel between parallel Universes---which thus are equally within the Sci-Fi realm.}.

\Sec{A Landau continuous Phase Transition: a Quantum Super-System}
 \Label{s:QSS}
Consider next a complex quantum system ``$1+2$'', or more precisely, a quantum super-system consisting of the quantum sub-systems $1$ and $2$. As standard, this super-system is equipped with a particular, correlationary, unitary quantum-dynamical evolution operator $\hU_{1+2}(t)$ and correspondingly, a correlated, quantum-dynamical state:
\begin{equation}
 \ket{\J^{1+2}\6(t)}=\hU_{1+2}\6(t)\ket{\J^{1+2}}=\sum_n c_n \ket{\J^1_n}\otimes \ket{\J^2_n\6(t)}
\Label{e:13}
\end{equation}
which evolves deterministically in time. Here, $\ket{\J^{1+2}}$ is the  initial quantum-dynamical state from the Hilbert superspace $\cH_1\otimes\cH_ {2qp}$ representing tensorial product of the Hilbert subspace $\cH_1$ and Hilbert's subspace $\cH_ {2qp}$ (on which the action of coordinate and momentum operators and analytic functions thereof is well-defined). Let $\cB_1=\{\ket{\J^1_n}, \forall n\}$ denote a time-\textit{independent} basis in $\cH_1$, and $\cB_2(t)=\{\ket{\J^2_n\6(t)}, \forall n\}$ a time-\textit{dependent} basis in $\cH_{2qp}$, $\cB_2=\{\ket{\J^2_n}, \forall n\}$ being the initial form of $\cB_2(t)$, while $c_n$, $\forall n$, are the \textit{constant} superposition coefficients that satisfy the unit norm condition analogous to~(\ref{e:2}). 

It is not hard to see that~(\ref{e:13}) possesses both a global $U(1)$ quantum-dynamical symmetry akin to~(\ref{e:3}), and also its sub-symmetry implemented by the unitary operator
\begin{equation}
 \hW_\ve[\J^{1+2}\6(t)] =
  \exp\left\{i\ve \ket{\J^{1+2}\6(t)}\bra{\J^{1+2}\6(t)}\right\}~,
\Label{e:14}
\end{equation}
with the continuous real parameter $\ve$, and the Lie group generator $\ket{\J^{1+2}\6(t)}\bra{\J^{1+2}\6(t)}$. 

Suppose again that there is such time $\tau$, so that when $t>\t$ then, for $\cB_2(t)$ and any observable $\hA_2$ that acts in $\cH_{2qp}$, certain approximation conditions analogous to $A_1$ and $A_2$ are satisfied. Also, let us point out that, after $t=\tau$, $\cB_2(t)$ is the unique (up to the $U(1)$ phase~(\ref{e:3}), \ie, including all of its globally $U(1)$-transformed pictures) basis in $\cH_{2qp}$ that is able to satisfy both approximation conditions. With these assumptions (to which we refer as the `sub-systemic weakly interfering wave packet approximation on the sub-system $2$' or, `sub-systemic approximation on $2$' for short), the state~(\ref{e:13}), under action of~(\ref{e:14}), turns into
\begin{equation}
 \wt\hW_\ve[\J^{1+2}\6(t)] \ket{\J^{1+2}\6(t)}= \sum_n c_n \exp\{i|c_n|^2\ve\}\ket{\J^1_n}\otimes \ket{\J^2_n\6(t)}~. 
\Label{e:15}
\end{equation}

The obtained quantum state~(\ref{e:15}) does not represent any global $U(1)$-transform of the quantum-dynamical sate~(\ref{e:13}) and so is not physically equivalent to it. That is, the corresponding superposition terms in~(\ref{e:13}) and~(\ref{e:15}) are \textit{decoherent}. Therefore, the given sub-systemic approximation on $2$ breaks the global $U(1)$ quantum-dynamical symmetry as well as the $\hW_\e[\J^{1+2}\6(t)]$ quantum-dynamical state symmetry (and prefers the unique basis $\cB_1{\otimes}\cB_2(t)$) on the whole super-system $1+2$. In other words, the given sub-systemic approximation on $2$ cannot be achieved by quantum-dynamical evolution~(\ref{e:13}) on the super-system $1+2$, so that the results of this sub-systemic approximation on $2$ must be discontinuously inaccessible by the exact quantum-dynamical evolution~(\ref{e:13}) on the quantum super-system $1+2$. 

This analysis, completely analogous to that in the previous section, points simply at the fact that the above sub-systemic approximation on $2$ produces a spontaneous superposition breaking on the super-system $1+2$, and a corresponding Landau continuous phase transition from $\ket{\J^{1+2}\6(t)}$ to some 
\begin{equation}
 \ket{\J^1_n}\otimes \ket{\J^2_n\6(t)}~,\qquad\mbox{for some}~n~,
\Label{e:16}
\end{equation}
with probability (statistical weight) of $w_n$, given analogously to~(\ref{e:12}). 

Owing to the correlation characteristics of the quantum-dynamical state~(\ref{e:13}), simultaneously to given sub-systemic approximation on $2$, the sub-system $1$ becomes described, with probability $w_n$, by the quantum state $\ket{\J^1_n}$ from uniquely (up to the global $U(1)$ phase) determined basis $\cB_1$ even if quantum states from $\cB_1$ satisfy no approximating condition. 

In this way, the sub-systemic approximation on $2$ effectively splits the super-system $1+2$ into $2$ (described by the corresponding statistical mixture of the quantum states from uniquely determined basis $\cB_2(t)$ whose quantum states satisfy given approximation conditions), and $1$ (described by correlated statistical mixture of the quantum states from the uniquely determined basis $\cB_1$ whose quantum states do not satisfy any approximation condition). 

But in the exact description (discontinuously distinct from the approximate sub-systemic description on $2$) of the quantum super-system $1+2$, this super-system is described exclusively by the \textit{correlated} quantum-dynamical state $\ket{\J^{1+2}\6(t)}$~(\ref{e:13}), which, within standard quantum-mechanical formalism\cite{6,8,9,10,11,14} and in full agreement with existing experimental facts\cite{22,23}, does not admit any separation of the quantum super-system $1+2$ into its quantum sub-systems $1$ and $2$ described by pure or mixed quantum states. Stated simply: the true state, $\ket{\J^{1+2}\6(t)}$, in general, does not factorize into anything like the sub-systemic approximation on $2$, $\ket{\J^1_n}\otimes \ket{\J^2_n\6(t)}$ for any $n$.

\Sec{Measurement As a Continuous Phase Transition}
 \Label{s:MeaSSB}
\sSec{The quantum measurement conundrum}
 \Label{s:QMC}
Let us first recall some facts about the quantum measurement. 

Consider a quantum system, the object of the measurement, $O$, which is, just before the measurement, exactly described by the quantum-dynamical state
\begin{equation}
 \ket{\J^O} = \sum_n c_n \ket{\J^O_n}~,
\Label{e:17}
\end{equation}
of unit norm and belonging to the Hilbert space $\cH_O$.
Let $\cB_O=\left\{\ket{\J^O_n}: \hA\ket{\J^O_n}=a_n\ket{\J^O_n}, \forall n\right\}$ be the eigen-basis of the time-independent measured observable, $\hA$, acting on $\cH_O$, and the $c_n$'s are constant superposition coefficients, which satisfy the normalization condition analogous to~(\ref{e:2}). 

During the short period of the measurement, $\ket{\J^O}$ turns exactly into some quantum state $\ket{\J^O_n}$ from $\cB_O$ with the probability $w_n$, given in a form analogous to~(\ref{e:12}). This transition is called the collapse (reduction, decoherence, \etc) of the state of the object.

Therefore, the collapse represents an exact (non-approximate) result of the measurement on $O$. It corresponds to an equally exact breaking of the quantum superposition~(\ref{e:17}). For this reason, the collapse cannot be explained by any quantum-dynamical evolution of the isolated object $O$, as any such evolution must be determined by a corresponding unitary operator that preserves superposition. This incompatibility is, essentially, the quantum measurement conundrum.

\sSec{Some attempts to resolve the conundrum}
 \Label{s:QMR}
The quantum-dynamical interaction between the object, $O$, and a measurement device, $M$, was modeled by von Neumann\cite{6} in following way. He supposed that, before the quantum-dynamical interaction with $O$ corresponding to the measurement,  $M$ is described by a quantum-dynamical state of unit norm, $\ket{\J^M_0}$, taken from $\cB_M=\{\ket{\J^M_n}, \forall n\}$, the eigen-basis of the so-caled pointer observable of $M$ on the Hilbert space $\cH_M$. Then, the quantum super-system $O+M$ is, before the quantum-dynamical interaction between $O$ and $M$ corresponding to the measurement, described by the uncorrelated quantum-dynamical state
\begin{equation}
 \ket{\J^O}\otimes \ket{\J^M_0}. 
\Label{e:18}
\end{equation}

Von Neumann represented a simplified, short-lived quantum-dynamical interaction between $O$ and $M$ by a practically time-independent unitary quantum-dynamical evolution operator, $\hU_{O+M}$, determined by the correlating condition:
\begin{equation}
 \hU_{O+M} \ket{\J^O_n}\otimes \ket{\J^M_m} = \ket{\J^O_n}\otimes \ket{\J^M_{n+m}}~,\qquad \forall n, m~.
\Label{e:19}
\end{equation}
Then, during the short-lived quantum-dynamical interaction between $O$ and $M$, the initial quantum state~(\ref{e:18}) evolves quantum-dynamically into the uniquely determined, correlated, final quantum state of $O+M$:
\begin{equation}
 \ket{\J^{O+M}} = \sum_n c_n\ket{\J^O_n}\otimes \ket{\J^M_n} . 
\Label{e:20}
\end{equation}

In addition, von Neumann\cite{6} supposed that the above experimental facts on the measurement on $O$ can be simply extended in the sense that the exact collapse on $O$ implies that there indeed exists an exact or absolute collapse, \ie, superposition braking on $O+M$. This is understood to mean that, after the measurement, $O+M$ is \textit{exactly} described by a quantum state
\begin{equation}
 \ket{\J^O_n}\otimes \ket{\J^M_n}
\Label{e:21}
\end{equation}
with probability $w_n$ given analogously to~(\ref{e:12}). However, it is very important to note that neither the supposition~(\ref{e:20}) nor the collapse of~(\ref{e:20}) into~(\ref{e:21}), have ever been verified experimentally so that this can only be regarded as a hypothesis, called the ``von Neumann projection postulate'' or the ``absolute collapse postulate.''

If one accepts the absolute collapse postulate, it follows immediately that the quantum-dynamical interaction between $O$ and $M$ described by~(\ref{e:20}) cannot explain this absolute collapse, since this quantum-dynamical interaction and absolute collapse are obviously discontinuously (inaccessibly) different. One must either reject the possibility of a complete dynamical description of the physical phenomena (by introducing the physically indescribable abstract Ego of the human observer\cite{6}, \etc) or extend the standard quantum-dynamical evolution. The latter is done through different types of non-unitary (nonlinear) types of dynamics with isolated $O+M$, \eg\cite{15}, by different types of non-unitary (nonlinear) dynamical interaction with environment\cite{16,17,18,19}, or finally, by introducing various types of hidden variables\cite{20}.

However, each of these groups of ``solutions'' is beset with problems\cite{7}. In the last case one considers, explicitly or implicitely, a non-spontaneous and dynamical breaking of the quantum superposition. Ref.\cite{15} attempts to build a ``unified dynamics for microscopic and macroscopic systems'' through a nonlinear dynamics of isolated $O+M$, which in turn causes a ``spontaneous localization'', \ie, a spontaneous collapse on $O$. Nevertheless, this ``spontaneous localization'' represents a dynamical unitary symmetry breaking completely different from the self-collapse as spontaneous superposition breaking in our sense. By contrast, in our proposal (see below), there is either the exact quantum dynamics or the discontinuously (inaccessibly) different approximate classical dynamics, and no unified quantum-classical dynamics. 

In Refs.\cite{16,17,18,19}, decoherence is a non-unitary dynamical interaction between the non-isolated $O+M$ and its environment, which causes an absolute transition from quantum to classical dynamics with dynamical breaking of the quantum superposition. A similar phenomenon of a dynamical breaking of the quantum superposition exists also in the different types of hidden variable theories\cite{20}. These are again completely different from the continuous phase transition from quantum to classical dynamics with spontaneous superposition breaking, as proposed herein.

Then, according to theoretical\cite{14} and experimental\cite{22,23} analyses, it follows that such supposed extensions of quantum dynamics cannot be local in sense of the theory of relativity, \ie, they must include some superluminal dynamical effects, which we find physically unjustifiable. That is, supposing that the standard quantum-mechanical formalism represents an averaging of a more precise dynamical formalism of sub-quantum, \ie, submicroscopic, mesoscopic, macroscopic or megascopic scales, it follows\cite{14,22,23} that such a more precise dynamical formalism must be superluminal. In particular, such nonlocal extensions of the quantum mechanics cannot be incorporated into contemporary relativistic quantum field theory\cite{5}.
 
The absolute collapse postulate may be rejected and replaced by a supposition regarding the relative collapse as it is phenomenologically (and in immediate agreement with experimental facts\cite{22,23}, but without a complete theoretical formalization) suggested in the Copenhagen interpretation\cite{10,11}. Here one supposes that only the correlated quantum-dynamical state $\ket{\J^{O+M}}$, as given in Eq.~(\ref{e:19}), describes the quantum super-system $O+M$ completely and exactly. That is, quantum mechanics represents an objective and complete theory of the super-system $O+M$, where the absolute collapse in form~(\ref{e:21}) does not actually occur. This supposition has also never been experimentally verified. Further, it is suggested that~(\ref{e:19}) may be approximated by~(\ref{e:20}), and this then can be called the relative (and \textit{effectively} exact) collapse on $O$ with respect to self-collapsed $M$, if $M$ is described effectively approximately, \ie, ``classical-dynamically'' in a phenomenological sense. Note that there exist different attempts\cite{25,16} of a consistent quantum formalization of the term ``classical-dynamically.'' However, they require only that $\cB_M(t)$ be a basis of weakly interfering wave packets, or, more precisely, that under a given approximating condition the state $\ket{\J^{O+M}}$, as given in Eq.~(\ref{e:20}), be transformed:
\begin{equation}
 \wt\hW_\ve[\J^{O+M}]\ket{\J^{O+M}}
  =\sum_n c_n \exp\{i \ve |c_n|^2\} \ket{\J^O_n} \otimes \ket{\J^M_n}~. 
\Label{e:22}
\end{equation}
Even as~(\ref{e:22}) is discontinuously (inaccessibly) distinct from~(\ref{e:20}), it is, as correctly pointed out in Ref.\cite{7}, just as distinct from~(\ref{e:21}), so that the above approximation is not sufficient for the complete quantum formalization of the relative collapse. 

\sSec{Measurement as a phase transition}
 \Label{s:MPT}
A simple comparison of the content of this and the previous two sections, \SS\,\ref{s:QSS} and \SS\,\ref{s:SQS}, indicates that the relative collapse could be unambiguously and completely modeled by the presented continuous phase transition from quantum into classical dynamics with spontaneous superposition breaking on the quantum super-system. Obviously, $O$ here corresponds to $1$, $M$ to $2$, and, the transition from~(\ref{e:20}) through~(\ref{e:22}) into~(\ref{e:21}) corresponds to the above phase transition with spontaneous superposition breaking from~(\ref{e:13}) through~(\ref{e:15}) into~(\ref{e:16}). Moreover since the presented superposition breaking is spontaneous and non-dynamical, it indicates that no superluminal effect (characteristic of dynamical breaking of superposition) should exist. All this would reaffirm and complete the formalization of the Copenhagen interpretation. 

A complete formalization of the relative collapse by a phase transition with spontaneous superposition breaking however still needs a generalization (entirely within standard quantum-mechanical formalism) of the simplified form of $\hU_{O+M}$~(\ref{e:19}). This generalization should correspond to $\hU_{1+2}(t)$ satisfying both~(\ref{e:13}) and~(\ref{e:19}) for $\cB_M$, which thus becomes time-dependent just like $\cB_2(t)$. 

The required generalization of $\hU_{O+M}$, \ie, von Neumann's quantum-dynamical interaction between $O$ and $M$ ($1$ and $2$, in general), can be realized relatively simply in the following way. Since the correspondence between the integral (with the evolution operator) and the differential form of the quantum-dynamical evolution is well-known\cite{8,9}, we turn to generalizing von Neumann's dynamical interaction between $1$ and $2$ in the differential form. 

Let the Hamiltonian observable of $1+2$ be time-independent (so that $1+2$ represents a conservative system) and let it have the following form:
\begin{equation}
 \hH_{1+2} = \hH_1 \otimes \Ione + \Ione\otimes \hH_2 + \hV_1 \otimes \hV_2~.
\Label{e:23}
\end{equation}
Here, $\hH_1$ is the Hamiltonian of the isolated sub-system $1$, $\hH_2$ the Hamiltonian of the isolated sub-system $2$, and $\hV_1 \otimes \hV_2$ the potential energy observable pertaining to the interaction between $1$ and $2$; $\Ione$ is the (appropriate) unit operator. 

Suppose now that $\hA_1$ represents the measured observable on the sub-system $1$, and that $\hA_1$, $\hH_1$ and $\hV_1$ all commute with each other so that they have common eigen-basis $\cB_1$, which we further take to be time-independent. 

Schr\"{o}dinger's equation for 1+2 then has form:
\begin{equation}
 \hH_{1+2}\ket{\J^{1+2}}
  =i\hbar \Big({\rd\over\rd t}\otimes\Ione + \Ione\otimes{\rd\over\rd t}\Big) \ket{\J^{1+2}}~,\qquad \ket{\J^{1+2}}\in\cH=\cH_1{\otimes}\cH_2~.
\Label{e:24}
\end{equation}
Suppose furthermore that a final solution of~(\ref{e:24}) is given in the form of the correlated quantum state~(\ref{e:13}) under the initial condition analogous to~(\ref{e:18}), and suppose that $\cB_2(t)$ is a time-dependent basis for the sub-system $1$. Upon projecting along $\cH_1$, Eq.~(\ref{e:24}) becomes
\begin{equation}
 \Big[\hH_2 + v_{1n}\hV_2\Big]\ket{\J^2_n}
 = i\hbar {\rd\over\rd t} \ket{\J^2_n}~, \quad\mbox{where}\quad
   \hV_1\ket{\J^1_n}=v_{1n}\ket{\J^1_n}~,\quad \forall n~,
\Label{e:25}
\end{equation}
and where $\ket{\J^1_n}\in\cB_1$ while $\ket{\J^2_n}\in\cB_2(t)$.

 Assuming that $\hV_1$ has a non-degenerate spectrum of eigenvalues\ft{The non-degeneracy assumption is merely a technical simplification here.}, Eq.~(\ref{e:25}) represents a system of \textit{mutually independent} equations with a common initial condition analogous to~(\ref{e:18}). For this reason, and according to the mathematical foundations of the standard quantum-mechanical formalism\cite{6,8,9}, it becomes obvious that $\cB_2(t) =\{\ket{\J^2_n}, \forall n\}$ cannot represent a basis in $\cH_2$ at any (and especially not at the initial) time. Therefore, the state~(\ref{e:13}) could not have been be a solution of Eq.~(\ref{e:24}), at any time. 

Nevertheless, it is possible that all quantum states from $\cB_2(t)$ satisfy the wave packet approximation as well as that the initial condition, before the measurement and corresponding to~(\ref{e:18}), is satisfied:
\begin{equation}
 \vev{\J^2_n(0)|\hx_2|\J^2_n\6(0)}
= \vev{\J^2_0|\hx_2|\J^2_0},\quad \forall n. 
\Label{e:26}
\end{equation}

Furthermore, we will assume that $\hV_2$ is chosen in such a convenient way that the expression
\begin{equation}
 \left|\vev{\J^2_n|\hx_2|\J^2_n} - \vev{\J^2_m|\hx_2|\J^2_m}\right|,\qquad
  \forall n, m \neq n~,
\Label{e:27}
\end{equation}
is a monotonously increasing function (from its initial value of zero), while the expression
\begin{equation}
 \inv2\left(\triangle_{\J^2_n}\hx_2 + \triangle_{\J^2_m}\hx_2\right)~,
  \quad \forall n, m \neq n~,
\Label{e:28}
\end{equation}
remains well-nigh constant in time (which implies that all disipations of the wave packets are neglected). 
 Obviously, the expressions~(\ref{e:26}) may be treated as a series of ordering parameters while the expressions~(\ref{e:28}) define the corresponding series of their critical values in a Landau continuous phase transition. 

Note that in the moment $\tau_{nm}$ when the expression~(\ref{e:27}) becomes equal to~(\ref{e:28}), $\J^2_n$ becomes weakly interfering and effectively (in the sense of averaged value approximation) orthogonal to $\J^2_m$, for $\forall n, m \neq n$. Finally, there is a time,
\begin{equation}
 \tau \ge \tau _{nm}~, \qquad \forall n, m \neq n~,
\Label{e:29}
\end{equation}
when all quantum states from $\cB_2(t)$ become effectively weakly interfering and orthogonal. At this moment, $\cB_2(t)$ also becomes effectively (approximately) a complete basis in $\cH_2$. (It is not hard to see that $\tau$ is finite for any finite and complete basis, $\hat{\cB}_2(t)$, in a finite subspace $\hat{\cH}_2\subset\cH_2$, which corresponds to real/actual measurements.)

Moreover, at the time $\tau$, the state~(\ref{e:13}) becomes an effective solution of Eq.~(\ref{e:24}). Thus, on one hand, quantum-dynamical evolution restitutes the correlations between the sub-systems $1$ and $2$, or, simply speaking, extends the superposition from $1$ onto the complete system $1+2$. On the other hand, at the time $\tau$, within a sub-systemic approximation of the weakly interfering wave packets of $2$ and according to the previous discussion, there occurs a self-collapse on $2$ and a relative collapse on $1$. Thus, the quantum-dynamical interaction between $1$ and $2$ and the measurement which $2$ realizes on $1$ both occur simultaneously, and both as the corresponding phase transitions, but in the discontinuously (inaccessibly) different levels of the analysis. 

Therefore, we conclude that the measurement can be completely modeled by a Landau continuous phase transition with spontaneous superposition breaking on the quantum super-system. 

\sSec{A simple example}
 \Label{s:Ex}
Let us consider now a concrete, simple but significant example, of a measurement modeled by a Landau continuous phase transition with spontaneous symmetry breaking. That is, the well-known\cite{12,13} example of Stern-Gerlach's spin measurement will be considered from the aspect of the described phase transition. 

Following the foregoing discussion, let $1+2$ be a single (Ag) atom in a magnetic field, $B_z$, directed along the z-axis. (The $x$-coordinate from the above discussion here becomes the $z$-coordinate.) As is usually the case, let
\begin{equation}
 \b_z = \Big(\pd{B_z}{z}\Big) \sim 10^3\,\mbox{T/m}~.
\Label{e:30}
\end{equation}
Here, $\cH_1$ will represent the ``internal'', two-dimensional spin Hilbert space, while $\cH_{2qp}$ will stand for the ``external'', or orbital Hilbert space for $z$-coordinate. Then, before interaction with the magnetic field, $1+2$ is described by initial quantum state
\begin{equation}
 \ket{\J^{1+2}_{in}} = \Big(c_-\ket{-} + c_+\ket{+}\Big)\otimes\ket{\J^2_0}~. 
\Label{e:31}
\end{equation}
Here, $\cB_1=\{\ket{-}, \ket{+}\}$ represents the eigen-basis of the $z$-component of the spin observable, $\hS_z$, with eigenvalues $\{-\inv2\hbar, +\inv2\hbar\}$, while $c_-$ and $c_+$ represent the corresponding superposition coefficients which satisfy the normalization condition
\begin{equation}
 |c_-|^2 + |c_+|^2 = 1 . 
\Label{e:32}
\end{equation}
Also, $\ket{\J^2_0}$ represents the initial wave packet of the center of mass of the atom, so that:
\begin{equation}
 \vev{\J^2_0|\hz|\J^2_0} = 0~, \qquad
 \vev{\J^2_0|\hp_z|\J^2_0} = 0~. \Label{e:34}
\end{equation}

Further, the potential is given by
\begin{equation}
 \hV_1\otimes\hV_2
    =-\Big(2\frac{\m_{\sss B}}{m\hbar}\hS_z\Big)\otimes B_z~,
\Label{e:35}
\end{equation}
where $\m_{\sss B}\sim10^{-23}$\,Nm/T is Bohr's magneton. Also, we may select
\begin{equation}
 \ket{\J^{1+2}}
  =c_-\ket{-}\otimes\ket{\J^2_-} + c_+\ket{+}\otimes\ket{\J^2_+}~,
\Label{e:36}
\end{equation}
where $\cB_2=\{\J^2_-, \J^2_+\}$ is the eigen-basis of $\hS_z$. Then 
\begin{equation}
 \vev{\J^2_{\pm}|\hz|\J^2_{\pm}} = \pm\frac{\m_{\sss B}}{2m}\,\b_z\,t^2~,
\Label{e:37}
\end{equation}
where $m\sim10^{25}$kg represents the mass of the atom, and
\begin{equation}
 \vev{\J^2_{\pm}|\hp_z|\J^2_{\pm}} = \pm\m_{\sss B}\,\b_z\,t~. 
\Label{e:38}
\end{equation}

Finally, a rough estimate yields:
\begin{equation}
 \inv2\Big(\triangle_{\J^2_-}\hz + \triangle_{\J^2_+}\hz\Big)
  \define\delta_z\sim 10^{-9}\,\mbox{m}
\Label{e:39}
\end{equation}
while
\begin{equation}
 \left|\vev{\J^2_+|\hz|\J^2_+}-\vev{\J^2_-|\hz|\J^2_-}\right|
  =\frac{\m_{\sss B}}{m}\,\b_z\,t^2~,
\Label{e:40}
\end{equation}
so that, the quantity~(\ref{e:40}) becomes equal to critical distance~(\ref{e:39}) at the critical moment
\begin{equation}
 t=\tau_c = \sqrt{\frac{\delta_z\, m}{\m_{\sss B}\b_z}} \sim 10^{-7}\,\mbox{s}~.
\Label{e:41}
\end{equation}

This shows that, after the action of the magnetic field and before any detector action on $1+2$, the self-collapse on $2$ and the relative collapse on $1$ occur already in typical microscopic domains, under the corresponding approximation conditions. At the same time, \ie, after the action of the magnetic field but before any detector action on $1+2$, the collapse on $1+2$ does not occur at all and $1+2$ is exactly described by corresponding correlated quantum state~(\ref{e:36}), as it is well-known\cite{12}.

\sSec{To summarize the foregoing}
 \Label{s:Sig}
Upon the above analysis of the measurement process, which demonstrated that measurement may be completely formalized in terms of spontaneous superposition breaking, the following conclusion emerges naturally.

In contradistinction to the understanding in theories of absolute collapse and hidden variables, quantum mechanics is hereby shown not to be an essentially mechanistic theory (a discretuum theory with actions at a distance). Instead, it is shown to be a truly complete field theory (a continuum theory with local actions, see Sec.~\ref{s:ExDist}) over an appropriate Hilbert space (see appendix~\ref{s:SSS}), the classical mechanic characteristics of which effectively emerge through a phase transition involving a spontaneous superposition breaking (see also appendices~\ref{s:PROB} and~\ref{s:WIWP}), as exhibited, \eg, in the case of quantum measurement. This establishes, and entirely within the standard formalism of quantum mechanics, a clear relationship between classical mechanics, quantum mechanics and quantum field theory.

\Sec{Quantum Mechanics As a Local Theory}
 \Label{s:LocQM}
Following the foregoing discussion and in agreement with the standard understanding of quantum physics\cite{6,8,9}, it follows that quantum-dynamical evolution provides the unique, completely exact form of the change of the quantum state in time. Thus, a quantum state represents the real physical ontology of the quantum system, and it is a subset of observables acting on the Hilbert space of such states that represents the real physical space. (It is only under special approximating conditions that a quantum state admits a reduction to the  ontology of classical mechanical, and the abstract space of coordinate observables acting on the Hilbert space to the usual ``real'' space.) This means that, as it has been supposed in the Copenhagen interpretation\cite{10,11}, quantum mechanics represents a complete and objective physical theory, even if the measurement (modeled here by spontaneous superposition breaking on the quantum super-system) represents a hybrid description of the dynamical interaction. This hybrid has been shown to be effectively approximate and ``classical'' on $M$, while effectively and relatively exact, but quantum on $O$. 
 
 It is well known\cite{9} that the non-relativistic, unitary, quantum dynamics (Schr\"odinger's equation) generalizes straightforwardly and without loss of unitarity into the special-relativistic quantum dynamics (the Klein-Gordon and the Dirac equations). In this sense, the dynamics of quantum mechanics is local, \ie, it is not superluminal. We now turn to show that the measurement process (formalized through a spontaneous superposition breaking), similarly leads to no superluminal effect. Together with the previous conclusion, this will then imply that quantum mechanics is, all in all, a fully local (field) theory.

Let the quantum super-system $1{+}2$ be described by a correlated quantum-dynamical state of unit norm,
\begin{equation}
 \ket{\J^{1+2}}=\sum_n c_n \ket{\J^1_n} \otimes \ket{\J^2_n}~,
\Label{e:42}
\end{equation}
where the $c_n$'s are constant superposition coefficients that satisfy normalization condition analogous to~(\ref{e:2}). Here $\cB_1={\ket{\J^i_n}, \forall n}$ represents a basis in an appropriate Hilbert space, $\cH_i$, for $i=1, 2$, where $\cH_1$ and $\cH_2$ are mutually equivalent. 

Now let $\hA,\hB,\hC,\hD$ be some four observables such that the condition
\begin{equation}
 \left|\vev{\J^i_n|\ha|\J^i_n}\right|\geq1~,\qquad
 \mbox{for}~\ha=\hA,\hB,\hC,\hD~,\quad \mbox{and}~i=1,2
\end{equation}
holds exactly. Suppose further that, for $\ha=\hA,\hB,\hC,\hD$ and over $\cB_2$, the following approximation conditions, similar to $A_1$ and $A_2$ in \SS\,\ref{s:SSB}, are satisfied: 
\begin{enumerate}
 \item[$a_1$:] $|\bra{\J^2_n}\ha\ket{\J^2_n} \gg \triangle_{\J^2_n}\ha \equiv \sqrt{\bra{\J^2_n}\ha^2\ket{\J^2_n} - \bra{\J^2_n}\ha\ket{\J^2_n}^2}$, $\forall n$, 
and
 \item [$a_2$:] $\bra{\J^2_n}\ha\ket{\J^2_m} \simeq \bra{\J^2_n}\ha\ket{\J^2_n}\delta_{nm}$, $\forall n, m$.
\end{enumerate}

Applying these approximating conditions to $\ket{\J^{1+2}}$, similarly to the application of the sub-systemic approximation of weakly interfering wave packets on the sub-system $2$, yields:
\begin{eqnarray}
 &&\kern-100pt\vev{\J^{1+2}|\hA \otimes \hB|\J^{1+2}} -
  \vev{\J^{1+2}|\hA \otimes \hD|\J^{1+2}}\nn\\
 &=&
 \sum_n|c_n|^2 \vev{\J^1_n|\hA|\J^1_n}\vev{\J^2_n|\hB|\J^2_n} -
  \sum_n |c_n|^2\vev{\J^1_n|\hA|\J^1_n}\vev{\J^2_n|\hD|\J^2_n} \nn\\
 &=&
 \sum_n |c_n|^2\vev{\J^1_n|\hA|\J^1_n}\vev{\J^2_n|\hB|\J^2_n}
  \Big(1\pm\vev{\J^1_n|\hC|\J^1_n}\vev{\J^2_n|\hD|\J^2_n}\Big)\nn\\
 && -
 \sum_n |c_n|^2\vev{\J^1_n|\hA|\J^1_n}\vev{\J^2_n|\hD|\J^2_n}
  \Big(1\pm\vev{\J^1_n|\hC|\J^1_n}\vev{\J^2_n|\hB|\J^2_n}\Big)
\Label{e:44}
\end{eqnarray}
that is,
\begin{eqnarray}
 &&\kern-100pt\Big|\vev{\J^{1+2}|\hA\otimes \hB|\J^{1+2}}-
   \vev{\J^{1+2}|\hA \otimes \hD|\J^{1+2}}\Big|\nn\\
 &\leq&
 2\pm\Big(\vev{\J^{1+2}|\hC\otimes \hD|\J^{1+2}}+
       \vev{\J^{1+2}|\hC\otimes \hB|\J^{1+2}}\Big). 
\Label{e:45}
\end{eqnarray}

The resulting inequality~(\ref{e:45}) is analogous, in form, to Bell's inequality\cite{14}. Of course, the original Bell inequality refers to arbitrary four quantum-mechanical observables acting in an arbitrary Hilbert space, averaged by objective hidden variables. Instead, the inequality~(\ref{e:45}) refers to four special quantum observables and their quantum-mechanical average values, as well as $\cB_2$ and $\cH_2$, which must satisfy the particular approximating conditions. For this reason, the inequality~(\ref{e:45}) does not have the general implications to quantum mechanics in the way that Bell's inequalities (separating quantum mechanics from hidden variables theories in general).

 Recall now that in real experiments\cite{22,23}, the original Bell's inequality is violated, implying in turn that all hidden variables theories must be superluminal or nonlocal if they are to be objective. It is not hard to see that in the same situation, there is no consistent way to obtain the inequalities~(\ref{e:45}) in a corresponding approximation. In this way, standard quantum mechanical formalism, including measurement as a spontaneous superposition breaking (hiding) within well-defined aproximation conditions, is in an excellent agreement with the existing experimental facts\cite{22,23} regarding the roughly, intuitively and not very precisely termed ``quantum non-locality'', ``quantum distant correlations'' or ``quantum entanglement''. Namely, although exact conservation includes an independence of the correlated quantum states or general superposition in a quantum super-system from distances in the usual space, this conservation does not depend on any super-quantum or super-luminal dynamics, but it is a natural consequence of the symmetries of the dynamically local quantum mechanical dynamical evolution and state.

 Thus, quantum mechanics is a truly local (field) theory, in the specific sense used throughout the present article.

\Sec{Experimental Distinction Between Absolute and Relative Collapse}
 \Label{s:ExDist}
As note above, a direct experimental verification of the existence of the correlated quantum state $\ket{\J^{O+M}}$ of $O+M$, as described in Eq.~(\ref{e:20}), has never been realized for a typical macroscopic system $M$ (with an Avogadro's number of atoms or more), for reasons of extreme techincal difficulties. On one side, a direct measurement of $\ket{\J^{O+M}}$, as given in~(20), (if it exists) as a measured quantum system by some new (and external to $O+M$) measurement device, $MM$, is technically very hard. 

On the other hand, before such a new measurement is done, it is very hard to eliminate various quantum-dynamical interactions (realized by exchange of a single photon or phonon, for example) between $O+M$ and its quantum-mechanically described environment, $E$. Such interactions produce, in many special cases, the correlated quantum-dynamical state of $O+M+E$:
\begin{equation}
 \ket{\J^{O+M+E}} = \sum_n c_n \ket{\J^O_n}\otimes \ket{\J^M_n} \otimes \ket{\J^E_n}
\Label{e:46}
\end{equation}
where $\cB_E=\{\ket{\J^E_n}, \forall n\}$ is a (time-independent) basis in the Hilbert space $\cH_E$ of $E$. The quantum super-system $O{+}M{+}E$, described by the correlated quantum-dynamical state~(\ref{e:46}), cannot be  factorized (seaprated), within standard quantum-mechanical formalism\cite{6,8,9}, into its quantum sub-systems each described either by pure or by mixed quantum states, or more precisely, by corresponding mixtures of the first kind representable by statistical operators\cite{9,26}. However, as a formal sub-system of $O{+}M{+}E$, $O{+}M$ may be described by a mixed state of the second kind\ft{The precise technical and physical distinction between (the usual) mixtures of the first kind and those of the second kind may be found in Ref.\cite{26}.}:
\begin{equation}
 \hr^{O+M}{}'' = \sum_n \vev{\J^E_n|\J^{O+M+E}}\vev{\J^{O+M+E}|\J^E_n} = \sum_n |c_n|^2 \ket{\J^O_n}\bra{\J^O_n} \otimes \ket{\J^M_n}\bra{\J^M_n}
\Label{e:47}
\end{equation}
This mixture~(\ref{e:47}) has the identical form,  mathematically, to the mixture of the first kind obtained by absolute collapse (if it exists) by measurement of $M$ on $O$. But the physical meaning of this mixture of the second kind~(\ref{e:47}) is only conditional (it depends on the basis $\cB_E$) and indirect. That is, suppose that for the quantum super-system $O{+}M{+}E$, described by~(\ref{e:46}), a measurement device (including $MM$) realizes new, sub-systemic (considering only dynamical interaction with a sub-system) measurement on $O{+}M$ as a formal sub-system of $O{+}M{+}A$. In this case, the obtained results are effectively the same as if $O{+}M$, before this new measurement and representing an isolated quantum system, has been in a mixture of the first kind, mathematically identical to a mixture of the second kind~(\ref{e:47}).

 But of course (if absolute collapse does not exist), before this new measurement, $O{+}M$ has been only a formal quantum sub-system of the quantum super-system $O{+}M{+}E$ in the pure quantum state~(\ref{e:46}). That is, even if a quantum-dynamical interaction between $O{+}M$ and $E$ does not generate an absolute collapse on $O{+}M{+}E$ (as any suggested non-unitary dynamical interaction with the environment\cite{16,17,18,19} which could generate such an absolute collapse must be superluminal), any later, incomplete, sub-systemic measurement on $O{+}M$ as a sub-system of super-system $O{+}M{+}E$, on the account of the neglecting $E$, will effectively lead to the conclusion that the absolute collapse has been occurred already by the measurement that $M$ realized on $O$.

Thus, at the quantum-mechanical level of analysis and in general, the (sub-systemic) mixtures of the second kind are conditional, \ie, they are not unambiguously determined. This fact is the main reason for Everett's many world or relative state interpretation\cite{MWI} not being complete at the quantum-mechanical level of analysis. Completing Everett's interpretation requires such an extension of the standard quantum-mechanical formalism in which the mixtures of the second kind would become unconditional, \ie, absolute. However, in that case, the super-system clearly also becomes described by an absolute mixture of the first kind, equivalent to absolute collapse.ÊThus, completing Everett's interpretation produces some kind of superluminal hidden variables theory, where every branch of theÊmultiverse corresponds so a homogeneous sub-quantum sub-ensemble. This provides a clear distinction between Everett's relative state and our relative collapse interpretation.

Note that the supposition regarding the environmental absolute decoherence \vs\ standard quantum-mechanical formalism (with relative collapse as spontaneous superposition breaking) corresponds in a significant way to Mach's principle \vs\ Einstein's general theory of relativity. Namely, both environmental absolute decoherence and also Mach's principle need an instantaneous action at a distance, contrary to fundamental principles and concepts of the relativistic, and in particular quantum field theory\cite{5}. For this reason new significant experimental data\cite{27} on the sub-systemic decoherence by (thermal) interaction with environment do not represent any conclusive fact on the existence of the absolute environmental decoherence or super-systemic collapse.

In this way, and owing to the noted technical difficulties, there is no unambiguos experimental evidence (for now) for the existence of absolute or relative collapse on $O+M$. By contrast, on only the sub-system $O$, both types of theories (the one with absolute and the one with relative collapse) yield effectively the same consequences, and which are in full agreement with existing experimental facts. However, even if there is no realistic possibility for experimental distinction between theory of the absolute and theory of the relative collapse in macroscopic domains as yet, it is possible that such experimental distinction can be relatively simply realized in microscopic domains. 

That is, from the aspect of the relative collapse theory, $O$ and $M$ can be quantum objects from micro, meso, macro or mega domains. For this reason we can chose such a microscopic $O$ and a microscopic or (quasi)mesoscopic $M$ so that it is unambiguous that $O+M$ is described by a correlated quantum-dynamical state, \ie, that quantum superposition on $O+M$ exists. Also, $O+M$ will be chosen in such way that any further quantum-dynamical interaction with its environment $E$ can be effectively neglected. Then we can introduce such an approximate sub-systemic description and experimental treatement of $M$ corresponding to the presented selfcollapse (as a Landau continuous phase transition with spontaneous superposition breaking) on $M$. If then, unambiguously, relative collapse occurs on $O$ as an effective quantum phenomenon it means that relative collapse theory, or, precisely, measurement modeled by given Landau phase transition and spontaneous superposition breaking, is experimentally affirmed. 

As a concrete example of such micro or (quasi)macroscopic measurement we shall recall briefly the demonstration\cite{28} of the appearance of the self- and relative collapse in an experiment of the quantum superposition of a mirror suggested by Marshall {\it et al.}\cite{21}. Here, a single photon that propagates through a modified Michelson's interferometer can represent $O$. That is, the modified interferometer has one usual (macroscopic and fixed) mirror, and another one which is unusual: it is (quasi)macroscopic (with $\sim 10^{14}$ atoms, or, with a linear dimension $\sim 10\mu m$) and it is oscillating; this can represent $M$. The quantum-dynamical interaction (by means of phonons and high-finesse cavities) between $O$ and $M$ correlates (entangles) $O$ and $M$ into $O+M$ and decorrelates (disentangles) $O+M$ into $O$ and $M$ alternatingly, \ie, periodically in time. More precisely, $O$ and $M$ are correlated during time intervals $(nT+\inv2\tau, (n+1)T-\inv2\tau)$ for $n=0,1,2,\ldots$, where $T$ and $\tau \ll T$ represent some positive time constants. Also, $O$ and $M$ are decorrelated during time intervals $(nT-\inv2\tau, nT+\inv2\tau)$ for $n=1,2,3,\ldots$. During any decorrelation time interval, $M$ is decribed by a wave packet while $O$ is described by a quantum superposition of "up-down" and "left-right" trajectory. During any correlation time interval, $O+M$ is exactly described by a correlated quantum state. In the given experimental circumstances that include extremely low temperature (less than 2\,mK) practically any environmental influence on $O+M$ can be neglected so that no absolute collapse (decoherence) occurs $O+M$ at any time (including the correlation time intervals). 

However, during any correlation time interval, $M$ can be sub-systemically and effectively approximated by a mixture of the second kind, consisting of initial rest wave packet from one and a quantum superposition of two movable wave packets from other side. (This efective sub-systemic superposition of two movable wave packets of $M$ represents the primary aim of Marshal {\it et al.}\cite{21}.) In given experimental circumstances all three wave packets refering on $M$ satisfy the weak interference approximation condition, so that on $M$ the self-collapse can occur. This means that the given superposition of the two movable wave packets cannot be observed directly sub-systemically, \ie, observed in an effective approximation. However, on the basis of quantum correlations between $O$ and $M$ and in relation to the self-collapsed $M$, $O$ is described by a (decoherent) mixture of the second kind of "up-down" (corresponding to mixture of movable wave packets of $M$) and "left-right" (corresponding to rest wave packet of $M$) quantum states. It corresponds to the relative collapse that occurs on $O$ as a consequence of the selfcollapse on $M$ during the correlation time intervals. Such relative collapse, \ie, decoherence on $O$ can be simply tested by an additional detector of the interference (even if, of course, such additional detection breaks any further periodical alternation of correlation and decorrelation on $O+M$).

So, not unlike the remarkable Michelson-Morley interferometer experiment that affirmed Einstein's relativistic tenet of absence of the absolute space in (non-quantum) mechanics and field theory, the experiment of Marshall \textit{et al}.\ on the quantum superposition of a mirror (in fact included in a Michelson interferometer) would affirm Bohr's tenet\cite{10,11} of the absence of absolute colapse (decoherence) in quantum mechanics. It is a fascinating curiosity that both experiments use practically identical experimental circumstances (while roughly a century apart), and that they both negate \textit{absolute} concepts, \ie, concepts of absolute breaking of fundamental physical symmetries.

Finally, another possible and important experimental distinction between absolute and relative collapse may be served by the ``delayed-choice for entanglement swapping'' (DCES)\cite{Rf1,Rf2} (once it is experimentally realized) to which we now turn, deferring a more detailed analysis for later. The original explanation of DCES\cite{Rf1,Rf2}, which represents one of the types of empirist-pragmatist interpretations of quantum mechanics (analyzing sub-systemic measurements on correlated systems performed by varions measuring devices), insists on the non-existence of any objective (individual) interpretation of quantum states, and that a ``quantum state is viewed as just a representative of information''\cite{Rf2}. Here is information \textit{ad hoc} understood, \ie, postulated as an essential category, characteristic to the inseparable information process (informational correlate or totality). Such a process includes both the source of information (the information object, \ie, the object, $O$, of the measurement) and the receptor of information (the information subject, \ie, the measuring device, $M$) but, by assumption, is not in any exact and unambiguous correspondence with any conceivable physical object. Clearly, this represents an extension of the usual quantum-mechanical formalism\cite{6,8} as well as the usual theory of information, and has been appropriately criticized\cite{SSN}.

 In our view, the DCES formalism is in full agreement with the standatd quantum-mechanical formalism and with the understanding of the quantum state of the super-system as a complete description of the quantum super-system as an unambiguous physical object. However, DCES clearly indicates the absence of the absolute collapse in the measurement process, and especially in sub-systemic measurement, which is confirmed in subsequent sub-systemic and super-systemic measurements. This also allows us to regard the exact unitary quantum-mechanical dynamical evolution (which correlates/entangles the sub-systems and the super-systems) or measurement as a reduced form of this evolution, as a realistic model of this information process, and which one must no longer postulate \textit{ad hoc}.

\Sec{Conclusions}
 \Label{s:Out}
We have shown that a type of a Landau continuous phase transition with spontaneous superposition breaking on the quantum super-system permits a complete and consistent formalization of the measurement within the unaltered standard quantum-mechanical formalism. This also confirms that the quantum-dynamical evolution represents the unique, completely exact change of the quantum state in time, so that quantum mechanics represents an objective and local physical theory.
 That is, quantum mechanics is a true field theory over an appropriate Hilbert space, the classical mechanic characteristics of which effectively emerge only through a phrase transition involving the spontaneous superposition breaking, \eg, during the measurement. This establishes the necessary and standard conceptual unity of classical mechanics, quantum mechanics and quantum field theory.
 Metaphorically speaking, and paraphrasing Bohr\cite{11}: The Good Lord (Nature) uses quantum dynamics and so needs no dice; but men do, to determine which classical r\^ole to take in the great drama of existence: an actor or a spectator, often not realizing that they are both.
 
\appendix
\Sec{On Probabilities}
 \Label{s:PROB}
 For the sake of completeness, we prove herein that the probability~(\ref{e:12}) given by spontaneous superposition breaking in the weakly interfering wave packet approximation really has the given form.
 
 Let a quantum system be described by a unit norm quantum state, $\ket{\J}\in\cH_{pq}$. Let $\cA\define\{\ket{a_n},\>\forall n\}$ be a basis of weakly interfering wave packets in $\cH_{pq}$. Then, owing to the weak interference of the wave packets approximation condition, the mathematically exact expression for the average value in the state $\ket{\J}$ of any observable $\hB$ that acts over $\cH_{pq}$ can be approximately presented by
\begin{equation}
 \Label{e:1*} \vev{\J|\hB|\J} \approx \sum_n \left|\vev{a_n|\J}\right|^2\vev{a_n|\hB|a_n}~,
\end{equation}
dropping the off-diagonal matrix elements, \ie, the interference terms.
                                              
 Let us choose any one particular term in the approximating expansion~(\ref{e:1*}):
\begin{equation}
 \Label{e:2*} \left|\vev{a_n|\J}\right|^2\vev{a_n|\hB|a_n}~.
\end{equation}
Of course, this corresponds to the local part, \ie, the superposition member
\begin{equation}
 \Label{e:3*} \vev{a_n|\J}\ket{a_n}~,
\end{equation}
for the correspondingly arbitrary $n$, of the expansion:
\begin{equation}
 \Label{e:4*} \ket{\J} = \sum_n \vev{a_n|\J}\ket{a_n}
\end{equation}
Since the norm of~(\ref{e:3*}) is, obviously, less than one in general, one may conclude that the approximate localization~(\ref{e:3*}) of $\ket{\J}$ and the corresponding approximate localization of~(\ref{e:2*}) of $\vev{\J|\hB|\J}$ do not have a direct physical meaning for any arbitrary $n$ as it violates the \textit{requirement of conservation of the unit norm for all reasonable physical quantum states}. Also, for the same reason, it would follow that $\vev{\J|\hB|\J}$ and $\ket{\J}$ cannot be consistently presented even by a simultaneous use of all their local parts, \ie, terms~(\ref{e:2*}) and~(\ref{e:3*}), respectively.
 
 However, the expression~(\ref{e:2*}) may be transformed into the equivalent expression
\begin{equation}
 \Label{e:5*} |\vev{a_n|\J}|^2\>
  {\vev{a_n|\hB|a_n}\over\triangle_{\ket{a_n}}\hB}
   ~~\triangle_{\ket{a_n}}\hB~,
\end{equation}
where, according to the characteristics of the approximate level of the analysis, $\vev{a_n|\hB|a_n}/\triangle_{\ket{a_n}}\hB$ may be consistently treated as the effective (``exact'') density distribution 
of the possible values of $\hB$ within the interval:
\begin{equation}
  I_{\ket{a_n}}(\hB) \define
  \Big(\vev{a_n|\hB|a_n}-\e_n^B,\vev{a_n|\hB|a_n}+\e_n^B\Big)~,\qquad
  \e_n^B\define\inv2\triangle_{\ket{a_n}}\hB~,
\end{equation}
for  arbitrary $n$.
  
 Now, one may suppose that~(\ref{e:5*}) can be given in the form
\begin{equation}
 \Label{e:6*} {\vev{a_n|\hB|a_n}\over\triangle_{\ket{a_n}}\hB}
   ~\triangle_{R\ket{a_n}}\hB~,
\end{equation}
where $\triangle_{R\ket{a_n}}\hB$ represents the width of the reduced interval, or subinterval:
\begin{equation}
 \Label{e:7*} I_{R\ket{a_n}}(\hB) = \Big(\vev{a_n|\hB|a_n}-\e_n^{RB},\vev{a_n|\hB|a_n}+\e_n^{RB}\Big)~,\qquad
   \e_n^{RB}\define\inv2\triangle_{R\ket{a_n}}\hB
\end{equation}
of $I_{\ket{a_n}}(\hB)$, for arbitrary $n$. This means that, in fact, an additional approximation is introduced. In this approximation, the local part, \ie, superposition term (of nonunit norm), $\vev{a_n|\J}\ket{a_n}$ of the complete (and unit norm) quantum state $\ket{\J}$ may be effectively presented by a new (and discontinuously different from $\ket{\J}$) proper (of unit norm) quantum state $\ket{a_n}$, for arbitrary $n$.
   
 However, such a representation can be valid only on the reduced subinterval $I_{R\ket{a_n}}(\hB)$ of the original, complete interval, $I_{\ket{a_n}}(\hB)$, of the possible values of $\hB$, for an arbitrary $n$. Then, the equivalence of~(\ref{e:5*}) and~(\ref{e:6*}) gives
\begin{equation}
 \Label{e:8*}\triangle_{R\ket{a_n}}\hB =
   |\vev{a_n|\J}|^2\triangle_{\ket{a_n}}\hB~,
\end{equation}
which can be treated as the effective definition (determination) of $\triangle_{R\ket{a_n}}\hB$, for all $n$. From this it follows that
\begin{equation}
 \Label{e:9*} |\vev{a_n|\J}|^2 =
    {\triangle_{R\ket{a_n}}\hB\over\triangle_{\ket{a_n}}\hB}~,\quad
\forall n~.
\end{equation}
Here, the expression $\triangle_{R\ket{a_n}}\hB$ may be treated as the measure (width or length) of the interval $I_{R\ket{a_n}}(\hB)$, while $\triangle_{\ket{a_n}}\hB$ may be understood to be the measure (width or length) of the original interval $I_{\ket{a_n}}\hB$, for all $n$. Then, according to the well-known ``geometrical'' probability definition\cite{Loev} and the given approximating conditions, Eq.~(\ref{e:9*}) can be treated as the probability that, within the given approximate analysis, a possible value of $\hB$ belongs to a conveniently reduced subinterval $I_{R\ket{a_n}}(\hB)$ of the complete interval $I_{\ket{a_n}}(\hB)$ of the values of $\hB$, for arbitrary $n$. Also, in this same approximation, $\ket{\J}$ may be effectively, probabilistically and globally represented by $\ket{a_n}$, for any arbitrary $n$.

In other words, if a quantum object formally and approximately treated as a classical particle should be found within the subinterval $I_{R\ket{a_n}}(\hx)$ of $I_{\ket{a_n}}(\hx)$ with the probability ${\triangle_{R\ket{a_n}}\hx\over\triangle_{\ket{a_n}}\hx}
 =|\vev{a_n|\J}|^2$, quantum superposition would be totally excluded, and the classical picture would be totally self-consistent and complete. Conversely, if this quantum object should be found outside the subinterval $I_{R\ket{a_n}}(\hx)$ but within $I_{\ket{a_n}}(\hx)$, it would follow that it cannot be represented (classically) by a wave packet.
   
Stated simply, by means of the probabilistic concepts within the given approximation of weakly interfering wave packets, a local representation of $\ket{\J}$ (which explicitly breaks the unit norm of the quantum state) can be formally, effectively, and discontinuously changed into a global representation of $\ket{\J}$ (of unit norm) by an arbitrary quantum state from $\cA$.
 
\Sec{The State Symmetry Structure over the Hilbert Space}
 \Label{s:SSS}
We are not aware of a detailed analysis in the existing literature of the symmetry structure to which we refer throughout this article, and will thus herein attempt to describe its basic characteristics. A full account is well outside the scope of this article, but we trust the Reader will understand the main gist of this intricate structure.

Consider a quantum-mechanical theory with an $N$-dimensional Hilbert space $\cH$. Being a vector space, it is of course possible to find many different bases for it, but let us specify a particular one, $\cB=\{\ket{u}_n:\vev{u_n|u_m}=\d_{n,m},\>\forall n,m\}$, where we will assume that $n,m$ range over a countable (finite, or infinite if $N=\aleph_0$) set (including a continuous (sub)range chiefly presenting notational and technical difficulties). An arbitrary state vector is then of course given by the familiar superposition $\ket{\j}=\sum_nc_n\ket{u_n}$. In \Eq{e:4}, we have defined a corresponding unitary operator (for notational simplicity we also ignore all time dependence at present),
 $$
 \hW_\ve[\j]=\exp\{i \ve \hP_\j\}~,\qquad
 \textrm{where}~~\hP_\j=\ket{\j}\bra{\j}~. \eqno(\ref{e:4}')
 $$
By acting (ultra-locally in the Hilbert space) on the state $\ket{J}$ itself, this operator merely transforms $\ket{\j}$ by a phase:
 $$
 \hW_\ve[\j]\,\ket{\j}=e^{i\ve}\ket{\j}~. \eqno(\ref{e:5}')
 $$
More generally, and owing to the idempotency of projection operators:
\begin{equation}
 (\hP_\j)^2\id\hP_\j~,\quad\To\quad
 \exp\{i \ve \hP_\j\}\id \Ione +(\e^{i\ve}{-}1)\hP_\j~,
\end{equation}
so that
\begin{equation}
 \exp\{i \ve \hP_\j\}\ket{\c}=\cases{
  \e^{i\ve}\ket{\c} &if $\ket{\c}\|\ket{\j}$, \ie,
                     if $\ket{\c}=c\ket{\j}$, for $c\in\IC$,\cr
   \noalign{\vglue3mm}
  \ket{\c} &if $\ket{\c}\perp\ket{\j}$, \ie,
            if $\vev{\c|\j}=0$,\cr
   \noalign{\vglue3mm}
  \ket{\c}+\vev{\j|\c}(e^{i\ve}{-}1)\ket{\j} &in general.\cr}
 \Label{e:BDef}
\end{equation}

Thus, the operator $\hW_\ve[u_n]$ acts (1)~as a $U(1)$ phase transformation on the basis vector $\ket{u_n}$ itself, but (2)~as the identity operator on all other basis elements $\ket{u_m}\in\cB$, for $m\neq n$. Consider then the family of such (basis-dependently defined) operators:
\begin{equation}
 \cW[\cB,\vec{\ve}\,]\define\{\hW_{\ve_n}[u_n],\forall n\}~,
\end{equation}
where $\vec{\ve}=(\ve_1,\ve_2,{\cdots})$ is the $N$-vector of transformation parameters. Recalling then the above-quoted two properties of these operators, \ie, the first two cases in the basic result~(\ref{e:BDef}), the family $\cW[\cB,\vec{\ve}\,]$ would seem to have the structure of a \textit{sheaf} (see Refs.\cite{Hirz,GrHa}) over the Hilbert space, $\cH$: to each basis element $\ket{u_n}$, $\cW[\cB,\vec{\ve}\,]$ associates an ultra-local copy of the abelian group $U(1)$---the \textit{stalk}, generated by the operator $\ket{u_n}\bra{u_n}$. In fact, we are now finally in the position to specify precisely: by ``ultra-local'' we imply that we ignore the algebro-geometric structure of the family $\cW[\cB,\vec{\ve}\,]$ not only globally over the whole Hilbert space $\cH$, but even in any arbitrarily small neighborhood of $\ket{\j}\in\cH$.

 Nevertheless, let us note that the two facts that: ({\it a})~$\cH$ is a vector space rather than just a topological space and ({\it b})~the third case in the basic result~(\ref{e:BDef}), jointly complicate matters considerably. The action of any one of the operators from the family $\cW[\cB,\vec{\ve}\,]$ on a general ``point'' in $\cH$ (represented by a general linear superposition of the $\ket{u_n}$) is a nontrivial combination of $U(1)$ phase transformations and ``translations'' (transforming superpositions) in $\cH$.

Consequently, the family $\cW[\cB,\vec{\ve}\,]$ constructed over the Hilbert space $\cH$ as above cannot be readily identified with any of the algebro-geometric structures well known and often used in theoretical physics, such as bundles and sheaves. $\cW[\cB,\vec{\ve}\,]$ is rather more complicated than that, although it does share some of the defining properties of these well-known and oft-used structures. Furthermore, it remains to carefully extract the basis-independent characteristics of the family $\cW[\cB,\vec{\ve}\,]$, constructed here in a manifestly basis-dependent fashion.

Finally, we note that the obvious quantum-dynamical relevance of the family $\cW[\cB,\vec{\ve}\,]$ provides a generalized gauge-theoretical structure to quantum mechanics. Rather importantly, the base space here is not the actual (real) spacetime as in the well-known gauge theories, but the Hilbert space. In view of this, the ultra-locality in the Hilbert space of the above definitions turns out to be most natural. Consequently, it is the nontrivial structure of the Hilbert space (it being both a vector space and being endowed with a convergent scalar product used to normalize the basis vectors) that induces the nontrivial action~(\ref{e:BDef}) of the operators~($\ref{e:4}'$) on any open neighborhood of $\ket{\J}\in\cH$, however small. In fact, the basic result~(\ref{e:BDef}) may easily be re-cast into the following infinitesimal ($\ve^2\ll\ve$) form:
\begin{equation}
 \Big[\Ione{-}\exp\{i\ve \hP_\j\}\Big]\ket{\c}\approx\cases{
  -i\ve\ket{\c} &if $\ket{\c}\|\ket{\j}$, \ie,
                     if $\ket{\c}=c\ket{\j}$, for $c\in\IC$,\cr
   \noalign{\vglue3mm}
  0 &if $\ket{\c}\perp\ket{\j}$, \ie,
                 if $\vev{\c|\j}=0$,\cr
   \noalign{\vglue3mm}
  -i\ve\vev{\j|\c}\,\ket{\j} &in general,\cr}
 \Label{e:BDDef}
\end{equation}
which allows us to interpret $[\Ione{-}\exp\{i\ve \hP_\j\}]$ as something like a covariant deformation operator on the family $\cW[\cB,\vec{\ve}\,]$. A not too dissimilar algebro-geometric structure is also found in the study of moduli spaces of Calabi-Yau manifolds (see Ref.\cite{Mrrsn,Beast} and the bibliography therein) where it leads to so-called variations of (mixed) Hodge structures. It is then tempting to conjecture that the analogue of the so-called Picard-Fuchs equation\cite{Mrrsn} from that study could play the r\^ole of the dynamical master equation in this generalized gauge-theoretic approach to quantum mechanics. A more precise formulation of this conjecture and its possible proof is however deferred to a later effort.

Fortunately, as the ultra-local properties exhibited above shall suffice for our present purpose, we defer a more careful global analysis of the family $\cW[\cB,\vec{\ve}\,]$ to a subsequent study.

For the rest of the discussion, we shall focus on a given initial state, $\ket{\J}=\sum_ic_i\ket{u_i}\in\cH$, and note that
\begin{equation}
 \hP_\J=\sum_i|c_i|^2\ket{u_i}\bra{u_i}
       +\sum_{j>i}c_i^*c_j\ket{u_j}\bra{u_i}
       +\sum_{j<i}c_i^*c_j\ket{u_j}\bra{u_i}~.
 \Label{e:Decomp}
\end{equation}
This corresponds to the well-known Gauss decomposition of unitary, $N{\times}N$ matrices into diagonal, upper- and lower-triangular matrices, and is also readily recognized in physics applications of group theory as the corresponding decomposition of generators of the $U(N)$ group into the diagonal (charge), raising and lowering operators.

Note that the operator~(\ref{e:4}), \ie,~($\ref{e:4}'$) associates a copy of this group structure to any given single state in the Hilbert space, and so also to the initial one, $\ket{\J}$. Let $G_\J$ denote this (non-abelian, $U(N)$-like) group defined ultra-locally at $\ket{\J}\in\cH$, and let $H_n$ denote the $U(1)$ subgroup generated by $\hP_{u_n}=\ket{u_n}\bra{u_n}$ for any one particular, fixed $n$.

Then, finally, the Goldstone modes discussed in the subsections~\ref{s:SSB}, \ref{s:DetSSB} and~\ref{s:Sum}, correspond to the ($SU(N)$-like) coset $G_\J/H_n$, here likewise defined ultra-locally. As seen from the decomposition~(\ref{e:Decomp}), the particular classicization-changing transformations discussed in the subsections~\ref{s:GM} and~\ref{s:NoGM} are indeed the raising and lowering generators of this ($SU(N)$-like) coset.

\Sec{Properties of Weakly Interfering Wave Packets}
 \Label{s:WIWP}
Let us provide a simple proof by contradiction that a nontrivial superposition of weakly interfering wave packets cannot itself, in general, be a wave packet.

 Suppose that a unit-norm but other wise arbitrary superposition, $\ket{\Psi}=\sum_nc_n\ket{u_n}$, of weakly interfering wave packets $\ket{u_n},\>\forall n$ and with coefficients $c_n,\>\forall n$, is itself a wave packet. Then, for every observable, $\hA$, over the appropriate Hilbert space, $\cH_{qp}$, it must be that
\begin{equation}
\Label{a1}
 \vev{\Psi|\hA^2|\Psi}\approx \vev{\Psi|\hA|\Psi}^2
\end{equation}
since, by assumption,
\begin{equation}
\label{a2}
 \vev{u_n|\hA^2|u_n}\approx\vev{u_n|\hA|u_n}^2~,\qquad \forall n~.
\end{equation}
However, as it is easily seen that expanding the left-hand side and one of the factors on the right-hand side of  Eq.~(\ref{a1}), the use of Eq.~(\ref{a2}) implies that:
\begin{equation}
\label{a3}
  \sum_n|c_n|^2\Vev{u_n|\hA|u_n}^2
  \approx\sum_n|c_n|^2\Vev{u_n|\hA|u_n}\Vev{\J|\hA|\J}~,
\end{equation}
and so, for a positive observable $\hA$,
\begin{equation}
  \vev{u_n|\hA|u_n}\approx\Vev{\Psi|\hA|\Psi} ~,\qquad
  \forall n~,
\end{equation}
which, in turn, violates the weak interference assumption~(\ref{e:9}).
This implies that that an arbitrary nontrivial superposition of weakly interfering wave packets cannot itself be a wave packet.

Furthermore, the wave packet (approximation) basis has the following straightforward property.

 Let $\hA(\hx)$ be an observable $\hA$ which is also an analytical function of the observable $\hx$. (In a special case, this observable may also represent a quantum-dynamical form, \eg, $\hA=\hH{-}i\hbar\pd{}{t}$. An expansion of the expression $\vev{\Psi|\hA(x)|\Psi}\define\vev{\hA\6(x)}$ into a Taylor series around $\vev{\Psi|\hx|\Psi}\define\vev{\hx}$ gives:
\begin{eqnarray}
\label{(1)}
 \vev{\hA\6(\hx)}&=&\Vev{\Big.\hA(\vev{\hx})}
          +\Vev{\Big.\hA'(\vev{\hx})}\vev{\hx-\vev{\hx}}
          +\inv2 \hA''(\vev{\hx})\Vev{(\hx-\vev{\hx})^2}
          +\cdots\nn\\
          &=& \vev{\hA(\vev{\hx})} + 0
            +\inv2 \hA''(\vev{\hx})(\triangle\hx)^2
            + \cdots
\end{eqnarray}
where $\hA'(\vev{\hx})$ and $\hA''(\vev{\hx})$ are the first and second derivatives of $\hA$ by its argument, $\vev{\hx}$.

 Roughly, the exact value on the right-hand-side of Eq.~(\ref{(1)}) may be expressed as a power series in $\triangle\hx$ (the standard deviation of $\hx$ in the given state), which very well corresponds to perturbation theory. Here, the $\triangle\hx$-independent term (the first term on the right-hand-side of~(\ref{(1)}) does not contain $\triangle\hx$. Similarly, the second term on the left of~(\ref{(1)}), which may be formally treated as a term linear in $\triangle\hx$, vanishes identically. Finally, it is only the third term on the right-hand-side of~(\ref{(1)}), \ie, the term quadratic in $\triangle\hx$, which is $\triangle\hx$-dependent and nonzero. The condition
\begin{equation}
\label{(2)}
\left|\vev{\hA\6(\hx)} - \Vev{\Big.\hA(\vev{\hx})}\right| \gg \inv2\left|\hA''(\vev{\hx})\right|(\triangle\hx)^2~,
\end{equation}
that is
\begin{equation}
\label{(3)}
 \triangle\hx \ll \sqrt{{|\vev{\hA\6(\hx)} - \hA(\vev{\hx})|\over
\inv2 \hA''(\vev{\hx}}}
\end{equation}
or, using~(\ref{(3)}),
\begin{equation}
\label{(4)}
   \vev{\hA\6(\hx)} \approx \left|\Vev{\Big.\hA(\vev{\hx})}\right|~,
\end{equation}
and
\begin{equation}
\label{(5)}
   \vev{\hA^2\6(\hx)} \approx \vev{\hA\6(\hx)}^2
\end{equation}
may then be regarded as the strict condition for the wave packet approximation. That is, any quantum state, $\ket{\J}$, which satisfies it may be regarded as a wave packet.

 Thus, in the wave packet approximation, the quantum dynamics (the left-hand-side of~(\ref{(1)}) for $\hA(\hx)$ as a quantum-dynamical expression) reduces to the first term on the right-hand-side of~(\ref{(1)}), which then represents the classical dynamics (with $\hA(\vev{\hx})$ as the dynamical expression) without any correction (terms containing $\triangle\hx$). Even for small, in the sense of Eqs.~(\ref{(2)})--(\ref{(5)}), indeterminacy in $\triangle\hx$, in the analysis of the coordinate $x$, the quantum dynamics may be, owing to the vanishing of the linear term, regarded as effectively equal to the classical dynamics.

 On the other hand, if one relaxes the conditions~(\ref{(3)}) and includes higher order (nonlinear) terms in the Taylor series~(\ref{(1)}), a corresponding correction of the classical dynamics is obtained, \ie, a semi-classical dynamics, which ultimately approaches the exact quantum dynamics.

 It is not hard to see that the satisfaction of the wave packet approximation conditions corresponds to the Heisenberg indeterminacy relations.

 Finally, when the wave packet approximation (and Heisenberg's indeterminacy relations) becomes violated, \ie, when the left-hand-side of~(\ref{(3)}) becomes equal to the right, the Taylor series~(\ref{(1)}) fails to converge. This simply means that the exact quantum dynamics, which of course continues to exist---the left-hand-side of~(\ref{(1)})---can no longer be consistently represented starting from the classical dynamics as the zeroth approximation. This also means that the Heisenberg indeterminacy relations specify the limits within which the quantum dynamics may be effectively projected into the classical.

\vspace {1.5cm} 
\noindent{\bf Acknowledgements}\hfil\break
The authors are very grateful to Prof.~Dr.~Fedor Herbut, whose original introduction of the  concept of relative colapse within hidden variables theories\cite{29,30,31,32} inspired this work. Also, authors are very grateful to Prof.~Dr.~Milan Vuji\v{c}i\'c and  Prof.~Dr.~Darko Kapor for illuminating discussions and support.
 T.~H.\ wishes to thank the US Department of Energy for their generous support under grant number DE-FG02-94ER-40854.


\end{document}